# Synthesis and Characterization of a LiFe$_{0.6}$Mn$_{0.4}$PO$_4$ Olivine Cathode for Application in a New Lithium Polymer Battery

Luca Minnetti[1], Vittorio Marangon[1], and Jusef Hassoun[1,2,3,*]

[1] *Graphene Labs, Istituto Italiano di Tecnologia, via Morego 30, Genova, 16163, Italy*

[2] *University of Ferrara, Department of Chemical, Pharmaceutical and Agricultural Sciences, Via Fossato di Mortara 17, 44121, Ferrara, Italy.*

[3] *National Interuniversity Consortium of Materials Science and Technology (INSTM) University of Ferrara Research Unit, University of Ferrara, Via Fossato di Mortara, 17, 44121, Ferrara, Italy.*

Corresponding Author: jusef.hassoun@iit.it, jusef.hassoun@unife.it

**Abstract**

A LiFe$_{0.6}$Mn$_{0.4}$PO$_4$ (LFMP) cathode exploiting the olivine structure is herein synthesized and characterized in terms of structure, morphology and electrochemical features in lithium cell. The material shows a relevant reversibility of the electrochemical process which evolves at 3.5 and 4 V vs. Li$^+$/Li due to the Fe$^{+2}$/Fe$^{+3}$ and Mn$^{+2}$/Mn$^{+3}$ redox couples, respectively, as determined by cyclic voltammetry. The LiFe$_{0.6}$Mn$_{0.4}$PO$_4$ has a well-defined olivine structure revealed by X-ray diffraction, a morphology consisting of submicron particle aggregated into micrometric clusters as indicated by scanning and transmission electron microscopy, and a carbon coating with a weight ratio of about 5% suggested by thermogravimetry. The electrode reveals in lithium cells subjected to galvanostatic cycling with conventional liquid electrolyte a maximum capacity of 130 mAh g$^{-1}$, satisfactory rate capability, excellent efficiency, and a stable trend. Therefore, the material is studied in a lithium metal polymer cell exploiting an electrolyte based on polyethylene glycol dimethyl ether (PEGDME) with the solid configuration. The cell reveals very promising features in terms of capacity, efficiency and retention, and suggests the LiFe$_{0.6}$Mn$_{0.4}$PO$_4$ material as a suitable electrode for a polymer battery characterized by increased energy density and remarkable safety.

**Table of Contents**

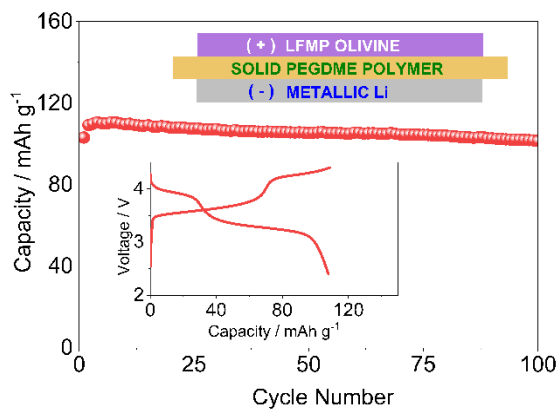

**A LiFe$_{0.6}$Mn$_{0.4}$PO$_4$ olivine** cathode operates between 3.5 and 4 V with a capacity over 130 mAh g$^{-1}$ and a remarkable cycle life. The electrode efficiently performs in a lithium metal polymer cell exploiting the PEGDME-SiO$_2$-LiTFSI-LiNO$_3$ solid electrolyte

**Introduction**

Lithium-ion battery (LIB) is the most diffused energy-storage system for application in popular electronic devices such as laptops and smartphones, and a very promising power source for electric vehicles (EVs) due to an energy density higher than 200 Wh kg$^{-1}$.[1,2] In this regard, the large scale diffusion of the latter EVs has been suggested to significantly limit the greenhouse gasses emission, and mitigate the atmosphere pollution, in particular in urban contests.[3] Hence, the LIB technology appeared a suitable approach for achieving a sustainable mobility, and partially mitigate the climate changes ascribed to the excessive global warming.[4,5] However, challenging targets of the market including long driving autonomy, high safety, and appealing price should be satisfied to actually boost the EVs and achieve the sustainable mobility.[6] Therefore, new configurations of LIBs have been developed over the last years with the aim of improving the performance and sustainability by adopting alternative chemistries at the anode, cathode and electrolyte sides.[7] Relevantly, new cathodes with higher voltage, improved capacity and remarkable stability with respect to the state-of-the-art appeared an adequate choice to achieve the next generation batteries.[8] In particular, phospho-olivines revealed remarkable stability as positive-electrode in rechargeable LIBs as well as additional bonuses including low cost, non-toxicity and modest environmental impact.[9] Among them, lithium iron phosphate (LiFePO$_4$) with a specific capacity of 170 mAh g$^{-1}$ and a redox potential of 3.5 V *vs* Li$^+$/Li has been commercialized since the last decade in LIBs with considerable thermal stability, lower cost, and higher safety content with respect to the conventional ones based on LiCoO$_2$ electrode.[7] The remarkable stability of the former electrode compared to the latter was attributed to the strong covalent bond of oxygen with phosphorus in the (PO$_4$)$^{3-}$ tetrahedra of the olivine poly-anionic framework with respect to the one of the layered oxide.[10–12] However, the lower potential of LiFePO$_4$ with respect to the conventional cathodes limits its energy density and suggests its possible use in less demanding fields such as standard-range urban electric vehicles or stationary energy storage.[13] A suitable strategy for increasing the energy density of electrodes with the olivine structure provides the use of transition metals with a higher redox potential with respect to iron, such as Mn, Co, or Ni into a "mixed-olivine" with the general formula LiFe$_{1-x}$M$_x$PO$_4$.[14,15] Therefore, high voltage mixed-olivine cathode for application in LIBs with improved performances may be achieved by exploiting the multi-metal concept,[15] whilst the presence of iron into the structure is suggested to increase the intrinsic electronic and ionic conductivity of the material,[16,17] and reduce the pseudo Jahn-Teller deformation of the other transition metals.[18] On the other hand, lithium metal anode may further improve the energy density of the battery; however, safety issues may be related to the possible formation of metal dendrites during lithium deposition/dissolution process.[19] Indeed, dendrites can cause short circuits of cells or battery sparks, thus leading to exothermic reactions of the flammable organic solvents in the electrolyte solution (e.g., highly volatile alkyl carbonates such as dimethyl-carbonate, DMC, and ethyl-carbonate, EC).[19–21] Therefore, the use of solid electrolytes exploiting amorphous polymers dissolving lithium salts, such as poly (ethylene oxide) (PEO),[22–24] or

poly(ethylene carbonate) (PEC),[25] has been suggested as the most satisfactory choice for achieving a safe use of lithium metal in battery. However, polymers such as PEO are typically employed with a relatively high molecular weight (MW ≥ 1000000 g mol$^{-1}$),[26] and have a poor ionic conductivity at temperature lower than 65 °C due to an excessive crystallinity hindering the Li$^+$ ions transport.[27] On the other hand, end-capped polyethylene glycol dimethyl ether (i.e., $CH_3$-$(OCH_2CH_2)_n$-$OCH_3$, PEGDME) with MW higher than 1000 g mol$^{-1}$ may be a suitable precursor for achieving solid polymer electrolytes operating at lower temperature than PEO.[28] In recent works we successfully reported safe lithium metal batteries using both $LiFePO_4$ and sulfur-carbon cathodes, benefitting from the solid polymer configuration ensured by a PEGDME with MW of 2000 g mol$^{-1}$.[29,30] The composite polymer electrolyte (PEGDME_CPE) revealed excellent properties in terms of Li$^+$ conductivity, thermal, chemical and electrochemical stability, as well as remarkable cycle life in solid polymer lithium battery operating at 50 °C.[29,30] Following the above described trends of electrodes and electrolytes for lithium battery, we reported herein the synthesis by sol-gel pathway and the full characterization of a $LiFe_{0.6}Mn_{0.4}PO_4$ olivine cathode with higher working potential and energy density compared to the common $LiFePO_4$ due to the presence of the manganese. The composition of the material developed herein is similar to the one previously investigated both in liquid and in polymer electrolyte i.e., $LiFe_{0.5}Mn_{0.5}PO_4$.[10,31] Hence, the above $LiFe_{0.5}Mn_{0.5}PO_4$ olivine cathode has shown at the room temperature in the conventional liquid electrolyte a capacity ranging from 120 mAh g$^{-1}$ at C/10 to about 85 mAh g$^{-1}$ at 1C,[10] whilst the $LiFe_{0.6}Mn_{0.4}PO_4$ developed in this work revealed a capacity ranging from about 130 mAh g$^{-1}$ at C/10 to 75 mAh g$^{-1}$ at 1C. On the other hand, the $LiMn_xFe_{1-x}PO_4$ may be further improved through the addition of new elements as dopants.[32] Indeed, $LiFe_{0.4}Mn_{0.595}Cr_{0.005}PO_4$/C prepared by ball milling/calcination revealed a specific capacity of 164 mAh g$^{-1}$ for 50 cycles at 0.1C,[33] $LiMn_{0.8}Fe_{0.19}Ni_{0.01}PO_4$/C synthesized by solvothermal/calcination process delivered 157 mAh g$^{-1}$ at 0.5C for 200 cycles,[34] $Li(Mn_{0.9}Fe_{0.1})_{0.95}Mg_{0.05}PO_4$/C prepared by mechano-chemical liquid-phase activation has shown 140 mAh g$^{-1}$ for 100 cycles at 1C,[35] $LiMn_{1/3}Fe_{1/3}V_{1/3}PO_4$/C achieved by ball milling/calcination delivered 122 mAh g$^{-1}$ for 100 cycles at 5C,[36] Li(Mn$_{0.85}$Fe$_{0.15}$)$_{0.92}$Ti$_{0.08}$PO$_4$/C obtained by ball milling/calcination performed 144 mAh g$^{-1}$ for 50 cycles at 1C,[37] LiMn$_{0.792}$Fe$_{0.198}$Mg$_{0.01}$PO$_4$/SGCNT synthesized by sol–gel/calcination delivered 119 mAh g$^{-1}$ for 3000 cycles at 1C,[38] LiMn$_{0.8}$Fe$_{0.19}$Mg$_{0.01}$PO$_4$/C prepared by ball milling/calcination revealed 109 mAh g$^{-1}$ at 10C,[39] LiMn$_{0.8}$Fe$_{0.19}$Mg$_{0.01}$PO$_4$/C synthesized by ball milling/calcination has shown 128 mAh g$^{-1}$ at 5C,[40] Li$_{0.995}$Nb$_{0.005}$Mn$_{0.85}$Fe$_{0.15}$PO$_4$/C prepared by ball milling/calcination revealed 146 mAh g$^{-1}$ for 50 cycles at 1C,[41] LiFe$_{0.48}$Mn$_{0.48}$Mg$_{0.04}$PO$_4$ performed 146 mAh g$^{-1}$ at 0.1C,[42] Li$_{0.97}$Na$_{0.03}$Mn$_{0.8}$Fe$_{0.2}$PO$_4$/C prepared by solvothermal/calcination delivered 125 mAh g$^{-1}$ for 200 cycles at 0.5C,[43] Li$_{0.98}$Na$_{0.02}$(Fe$_{0.65}$Mn$_{0.35}$)$_{0.97}$Mg$_{0.03}$PO$_4$/C prepared by sol–gel/calcination has shown 148 mAh g$^{-1}$ for 40 cycles at 0.1C,[44] LiFe$_{0.4}$Mn$_{0.6}$(PO$_4$)$_{0.985}$I$_{0.015}$ prepared by ball milling/calcination revealed 122 mAh g$^{-1}$ for 50 cycles at 1C,[45] and a LiFe$_{0.4}$Mn$_{0.6}$PO$_{3.97}$F$_{0.03}$ synthesized by ball milling/calcination delivered 153 mAh g$^{-1}$ at 0.1C.[46] However, we point out that the simplicity of our approach may actually favor the scaling up of commercial olivine cathodes with enhanced properties compared to those presently available in the market.

The synthesis of the LiFe$_{0.6}$Mn$_{0.4}$PO$_4$ material is advantageously monitored by thermogravimetric analysis (TGA) and X-ray diffraction (XRD), while the material is studied in terms of structure, thermal, morphological and electrochemical features in lithium cell with conventional electrolyte. Furthermore, we originally investigated the new electrode in a lithium cell exploiting the solid polymer configuration of the PEGDME-based composite electrolyte, operating at 50 °C with two voltage plateaus centered at about 3.5 and 4 V. The achieved results may possibly pave the way for a new generation of lithium polymer cells characterized by high energy, remarkable stability and, at the same time, suitable safety and environmental characteristics.

**Results and Discussion**

The synthesis of the LiFe$_{0.6}$Mn$_{0.4}$PO$_4$ proceeds with a series of annealing steps under argon of solid precursors obtained by sol-gel precipitation (see experimental section for details). Figure 1 depicts the related weight change monitored by TGA during the heating processes (panel a), and the various isothermal steps (panels b-e), as well as the corresponding modification of the precursor structure detected trough XRD (panels f-i). The TGA of Fig. 1a reveals a first extended weight loss occurring between 150 and 350 °C, likely ascribed to the degradation of the citric acid precursor with formation of methyl maleic anhydride, water and carbon dioxide (see left-hand reaction inset).[47,48] A second weight loss with a lower magnitude compared to the first occurs at temperatures higher than 450 °C, and may be ascribed to the degradation of methyl maleic anhydride to form water, carbon dioxide and elemental carbon (see right-hand reaction inset in Fig. 1a).[49] The TGA indicates an overall weight decrease of the precursor extending to about 40 % upon the whole annealing process, to finally achieve the thermally stable LFMP olivine cathode.[50,51] Furthermore, isotherms at 350 °C (Fig. 1b), 600 °C (Fig. 1c), 700 °C (Fig. 1d) and 850 °C (Fig. 1e) employed to achieve the steady state at the end of each annealing step reveal significant weight loss only at 350 °C, thus accounting for the stability of the olivine, which is formed already after at 700 °C.[52] Indeed, XRD patterns are collected after each synthesis step to evaluate the evolution of the crystalline phase. The XRD of the dry gel (Fig. 1f) and of the precursor upon 350 °C (Fig. 1g) show the expected amorphous structure, while the crystalline phase growths at 700 °C (LFMP_700, Fig. 1h) as above mentioned, and the final cathode material with the well-defined peaks characteristic of single-phase olivine structure (ICSD #193641 indexed in Figure S1a in Supporting information) is achieved upon 850 °C (LFMP_850, Fig. 1i).[53]

**Figure 1**

Figure 2 reports the detailed structural investigation by Rietveld refinement of the diffractograms related to the intermediate olivine material achieved at 700 °C (LFMP_700, panel a) and of the final one obtained at 850 °C (LFMP_850, panel c), as well as the TGA under air of the samples performed to calculate the carbon content of the two materials (panels b and d, respectively). The experimental XRD patterns of Fig. 2a and Fig. 2c (black dots) show that the both LFMP_700 and LFMP_850 are characterized by an orthorhombic cell unit (*Pnma* space group, table N. 62) without significant content of impurities, according to the reference pattern ICSD #193641 of the olivine lattice, as indeed shown by the refined diffractograms (red line) and the difference profile (blue line).[54] The refinement results reported in Table 1 reveal for the two samples similar values of the cell unit parameters *a*, *b* and *c*, that is, of about 10.4 Å, 6.0 Å, and 4.7 Å, respectively[9,54,55]. Therefore, the samples synthesized herein have a cell volume of about 294.9 Å$^3$ for the LFMP_700 intermediate and of about 295.8 Å$^3$ for LFMP_850, which exceed that of the ICSD reference reported in Table 1 due to the higher manganese fraction achieved in the LFMP_700 and LFMP_850 olivine.[56] Table 1 also reveals an increase of the crystallite size from 95 nm in LFMP_700 to 147 nm in LFMP_850, as most likely due to the expected coalescence of the crystalline domains during the thermal treatment at 850 °C.[57,58] This trend is confirmed by the application of the Scherrer equation to the main peak (2θ = 35.4°) of the two diffractograms as reported in Figure S1b and S1c in the Supporting information.[59] Despite the increase of the crystallite size may partially hinder the charge transfer, it can actually favor the reversibility of the material, and strongly mitigate side reactions with the electrolyte particularly occurring at the higher potential values.[60] In addition, Table 1 shows values ≤ 2 and ≤ 15 for the good-of-fit parameter (GOF) and weighted-profile R factor ($R_{wp}$%) of LFMP_700 and LFMP_850, respectively, that can be considered suitable and point out the reliability of the refinement.[61] Furthermore, the well-defined structure, the absence of excessive deformations, and the negligible impurity content represent key factors for allowing the efficient operation of the LiFe$_{0.6}$Mn$_{0.4}$PO$_4$ in lithium battery.[60] Another important aspect for this matter is represented by the carbon content in the material: indeed, excessive presence of this electrochemically inactive element may jeopardize the energy density of the material, instead a too low carbon amount can hinder the reaction kinetics of the intrinsically insulating olivine.[62] Hence, a carbon content ranging from 3 to 5 % can actually allow the most suitable performance of the olivine electrode in lithium cell.[63] This key parameter is herein determined by analyzing the TGA curves from 25 to 800 °C under air in Fig. 2b (LFMP_700) and Fig. 2d (LFMP_850), in which the weight changes are ascribed to the oxidation of the olivine materials to form transition metal oxides, phosphates, lithium oxide, and carbon dioxide as illustrated by reaction (1) in the experimental section.[64,65] Therefore, according to equations (2) and (3) (experimental section) the carbon weight ratio in LFMP_700 and LFMP_850 is calculated to be of about 6 % and 4 %, respectively. These results suggest optimized synthesis conditions leading to a slight excess of carbon at 700 °C, which is decreased by annealing at 850 °C to fully achieve the ideal target allowing the best performance of the $LiFe_{0.6}Mn_{0.4}PO_4$ olivine material in lithium cell.[65,66]

**Figure 2**

Figure 3 shows the morphological features of final $LiFe_{0.6}Mn_{0.4}PO_4$ sample (LFMP_850), investigated by scanning electron microscopy (SEM), energy dispersive X-ray spectroscopy (EDS) and transmission electron microscopy (TEM). The related SEM images at various magnification indicate the presence of agglomerates with a size of about 30 µm (Fig. 3a), formed by smaller primary particles (Fig. 3b) with a submicron diameter (Fig. 3c) ranging from 300 to 900 nm. This range is fully confirmed by the analysis of the SEM image aimed to determine the particle size distribution reported in Figure S2 in the Supporting Information section, with a maximum centered at about 500 nm. This composite morphology can ensure at the same time a short diffusion pathway for the lithium ions, and limited side reactions with the electrolyte. Indeed, particles with dimensions of the order of few hundreds nm allow fast kinetics, while micrometric agglomerates with limited surface area hinder the electrolyte decomposition, particularly at high voltage values.[63] Furthermore, the EDS mapping of C (Fig. 3d), Fe (Fig. 3e), Mn (Fig. 3f), P (Fig. 3g), and O (Fig. 3h) of the material surface detected by SEM suggests a homogeneous distribution of the various elements, and excludes the presence of impurities, in agreement with XRD of Fig. 2. Relevantly, the EDS spectrum of LFMP_850 reported in Figure S3 (Supporting Information) approximately confirms the target stoichiometry of the $LiFe_{0.6}Mn_{0.4}PO_4$, with an elemental ratio of Fe and Mn of 0.57 and 0.43, respectively. This result suggests the full precipitation of the metal precursors during the initial sol-gel step which is a key factor to achieve the proper olivine structure, and the optimal electrochemical activity of the material. Moreover, the TEM images (Fig. 3i-k) confirm the submicrometric size of the primary particles, and evidence the predominance of a spherule-shaped morphology. Furthermore, the TEM reveals the presence of a thin and amorphous material interconnecting the various LFMP_850 domains, which likely represents the conductive carbonaceous matrix detected by TGA (Fig. 2) deriving from the degradation of the citric acid precursor during synthesis (see complete reaction in Figure 1a). In this regard, the carbon revealed by TEM in the LFMP_850 olivine and quantified by TGA (about 5 %) appears as a layer extending from about 50 nm to 100 nm within the active material particles rather than an homogeneous carbon coating. In summary, both structural and morphological features of the mixed olivine synthesized in this work indicate its possible application in efficient energy storage devices characterized by a well reversible electrochemical process.[67,68]

**Figure 3**

The electrochemical process of the LFMP_850 cathode is subsequently investigated in lithium cell by combining cyclic voltammetry (CV) and electrochemical impedance spectroscopy (EIS), as reported in Figure 4. The CV response shown in Fig. 4a indicates the occurrence of two oxidation processes revealed by peaks centered at about 3.55 V vs. $Li^+/Li$ and 4.1 V vs. $Li^+/Li$ during the anodic scan, which are almost completely reversed into two reduction processes with peaks at about 3.45 V and 3.9 V vs. $Li^+/Li$ during cathodic scan. Indeed, the peaks at 3.55 and 3.45 vs. $Li^+/Li$ are ascribed to the $Fe^{+2}/Fe^{+3}$ redox couple whilst those at 4.1 and 3.9 V vs. $Li^+/Li$ to the $Mn^{+2}/Mn^{+3}$ one, with de-insertion of $Li^+$ from LFMP_850 during the anodic scan, and its insertion back into the olivine structure during the cathodic scan.[10] It is worth mentioning that the electrochemical process attributed to the $Fe^{+2}/Fe^{+3}$ redox couple in the LFMP_850 occurs at higher potential compared to bare $LiFePO_4$, while that of $Mn^{+2}/Mn^{+3}$ evolves at lower values with respect to $LiMnPO_4$.[66,69] The shift of the transition metal redox potentials is typically observed in mixed olivine materials, and attributed to changes in the ionic character and distance of the Me – O bond by substitution.[31,70,71] In particular, a modification in the covalency of the Me-O bond and Me-O-Me interaction has been reported to modify the position of the $Me^{+2}/Me^{+3}$ redox energy, thus directly influencing the corresponding redox potential.[71] The CV profile suggests a remarkable reversibility, modest polarization and a relatively fast kinetics of the electrochemical $Li^+$ (de)insertion process for the LFMP_850 electrode.[72] Furthermore, the first cycle of Fig. 4a (black line) reveals a slightly higher polarization than the subsequent ones (blue lines), thus indicating an activation process likely attributed to a partial structural reorganization of the material, as well as to the formation of a suitable solid electrolyte interphase (SEI) at the electrode surface.[73] To further clarify this aspect, Fig. 4b shows the EIS measurements at the open circuit condition (OCV) of the cell as well as after 1, 5 and 10 CV scans, and Table 2 reports the interphase resistances obtained by non-linear least squares (NLLS) analysis.[74,75] The Nyquist plots generally reflect the blocking-electrode shape expected by the fully-discharged state of the cell (i.e., at about 3 V vs. $Li^+/Li$ at the OCV, and at 2 V vs. $Li^+/Li$ after the CV scans) upon which the EIS is collected, which is characterized by the presence of a medium-high frequency semicircle accounting for the interface, and a low-frequency tilted line representing the cell geometrical capacity.[76] Therefore, the EIS at the above mentioned state of charge may be represented by the equivalent circuit $R_e(R_iQ_i)Q_g$, including *in series* the electrolyte resistance ($R_e$), one or two constant phase/resistance elements ($R_iQ_i$) mainly accounting for the SEI formed at the electrodes, and a final constant phase element related the cell capacitance ($Q_g$).[74,75] The data of Table 2 indicate low overall values of the interphase resistance ($R_i$), which decreases from about 19.5 Ω at the OCV to about 7.5 Ω after 10 voltametric cycles, thus accounting for the formation of an optimal SEI as well as for the above mentioned structural reorganization, which favors the reaction kinetics and improves the electrode/electrolyte interphase in the lithium cell.[77]

**Figure 4**

The electrochemical performances of LFMP_850 in lithium cell subjected to galvanostatic cycling tests are shown in Figure 5. The rate capability of the material is evaluated in terms of the voltage profiles (Fig. 5a) and the cycling trend (Fig. 5b) by increasing the current from C/10 to 2C (1C = 170 mA g$^{-1}$). The cell reveals the expected shape of the voltage profiles at the various C-rates (Fig. 5a), characterized by two plateaus at an average voltage of 3.5 V and 4 V, ascribed to the redox couples $Fe^{2+}/Fe^{3+}$ and $Mn^{2+}/Mn^{3+}$, respectively, in agreement the CV of Fig. 4a. Furthermore, the profiles of Fig. 5a suggest an increase of polarization by raising the current which leads to the decrease of the capacity from 130 mAh g$^{-1}$ (i.e., about 77 % of the theoretical value) at C/10 to about 65 mAh g$^{-1}$ (39 % of the theoretical value) at 2C, as indeed expected by the increase of the ohmic polarization. On the other hand, the cycling trend of Fig. 5b shows that the initial capacity of about 130 mAh g$^{-1}$ is recovered by decreasing the current back to C/10 after the whole test, thus suggesting a good rate capability and the stability of the LFMP_850 cathode upon the stress caused by raising the currents.[78] The stability of the electrochemical lithium (de)insertion process of LFMP_850 in cell is further investigated by repeated charging/discharging at a constant C-rate. The voltage profiles of the test performed at C/10 (Fig. 5c) and C/5 (Fig. 5e) show the well-defined plateaus above described, and a maximum capacity of about 130 mAh g$^{-1}$ at C/10 and 120 mAh g$^{-1}$ at C/5, in full agreement with the rate capability test. The cell at C/10 reveals a coulombic efficiency of about 90 % at the first cycle which increases and ranges from about 97 to 99 % during the subsequent ones, with a capacity retention of about 95 % (Fig. 5d). Instead, the cell at C/5 shows an initial coulombic efficiency of about 85 % increasing to a about 100 % during the subsequent cycles, and a capacity retention of about 90 % (Fig. 5f). It is worth mentioning that the galvanostatic cycling tests reported in Figure 5 are carried out in T-type cells, which allow the evaluation of the best performance of the LFMP_850 material in terms of the delivered capacity. Further cycling tests extended over 300 cycles are performed at the higher C-rate of 1C in coin-type cell (see experimental section) in order to evaluate the cycle life of the LFMP_850. The results reported in Figure S4 (Supporting Information) reveal at the end of the test a voltage signature without significant modifications compared to the steady state (Fig. S4a), a coulombic efficiency approaching 100% after the first few cycles, and a very stable capacity trend (Fig. S4b). Indeed, the cell retains the 85 % of the initial capacity over the 300 cycles investigated herein. Relevantly, the cycling test has very high coulombic efficiency, remarkable capacity and excellent retention, thus suggesting the high stability of the sample as expected by the optimal olivine structure.[79–81] However, a capacity value lower than the theoretical one (i.e., of approx. 80 % at C/10 and 45 % at 1C), and a lower rate capability than that of commercial LiFePO$_4$ suggests the need for a further improvement of the synthetic steps.[14] On the other hand, we remark that the presence of a second voltage step at about 4 V in LiFe$_{0.6}$Mn$_{0.4}$PO$_4$ rather than the only one at 3.5 typical of the LiFePO$_4$ actually represent a bonus for increasing the energy density of the material in lithium cell.[9,66]

**Figure 5**

The structural and morphological retention of the LFMP_850 electrode upon prolonged cycling in lithium cell has been evaluated in Figure 6. Indeed, the panel a of this figure displays the cycling trend related to the above galvanostatic test carried out at the constant current rate of C/5 for over 250 charge/discharge cycles (inset shows a related steady state voltage profile), and reveals an outstanding capacity retention approaching 90 %. After the cycling test, the cell is disassembled and the electrode recovered and compared with a pristine LFMP_850 electrode through XRD analyses (Fig. 6b). Hence, the XRD patterns of the pristine and cycled electrode exhibit a defined crystalline structure, in full agreement with the one observed in Fig. 1i for the LFMP_850 powder, thus indicating the retention of the single-phase olivine structure and evidencing the notable stability of the material upon prolonged operation in lithium cell. On the other hand, the morphology of the cycled LFMP_850 electrode is studied by SEM and compared with a pristine one, as reported in Fig. 6c-f. The pristine electrode (Fig. 6c and d) shows the expected submicrometric LFMP_850 primary particles aggregated into micrometric secondary particle and homogeneously blend into the electrode film (see experimental section in the Manuscript for details on electrode preparation). The particles evidence almost unaltered aspect in the cycled electrode (Fig. 6e and f) without relevant reorganization of the LFMP_850 domains despite long cycling in lithium cell. Furthermore, Fig. 6e and f reveal for the cycled electrode the presence of domains with rough morphology dispersed within the primary particles and on the electrode film, which are not observed in the pristine sample (compare Fig. 6c,d and Fig. 6e,f), as well as dispersed glass fibers due to the electrolyte separator (see experimental section for cell assembly details). The additional domains are likely associated with the formation of a SEI passivation layer during cycling as observed in previous work.[82] Therefore, the exceptional stability of both structure and morphology of the LFMP_850 electrode further suggest the suitability of this material for application in a novel configuration of lithium cells exploiting olivine-based cathodes operating at raised voltages.

**Figure 6**

Certainly, the presence of a voltage step over 4 V due to the $Mn^{+2}/Mn^{+3}$ redox couple represents a remarkable advantage of $LiFe_{0.6}Mn_{0.4}PO_4$ olivine cathode, and the use of the light lithium metal is a further bonus in terms of energy density. However, the use of this intriguing anode with a liquid electrolyte, such as that typically employed for material characterization (herein EC:DMC, $LiPF_6$), is actually hindered by above mentioned safety reasons related with the flammability of carbonate-based solutions, and possible formation of dendritic structures at the metal surface.[19–21] On the other hand, polymer solid-solutions based on PEO may allow the safe use of the Li-metal in practical cells; however, the latter electrolyte operates typically at temperature higher than 65 °C and has in this condition an anodic stability limited to about 4 V vs. $Li^+/Li$.[22–24] Differently, the PEGDME_CPE can operate at 50 °C with anodic stability extending at least over 4.4 V vs. $Li^+/Li$, as demonstrated by our early paper.[29] Therefore, we have originally used this alternative electrolyte in a safe lithium polymer cell using the LFMP_850 electrode investigated above. Figure 7 reports the electrochemical performances in terms of galvanostatic voltage profiles (panel a) and cycling trend (panel b) of the Li|PEGDME_CPE|LFMP_850 polymer cell at 50 °C using a current rate of C/5, and a voltage ranging from 2.4 to 4.4 V. Prior to cycling, the cell is heated at 70 °C overnight to achieve a predominantly amorphous state of the polymer, and improve the electrode/electrolyte interphase as demonstrated by the EIS analysis reported in Figure S5 and Table S1 (Supporting Information), that reveal an overall resistance of about 273 $\Omega$ and suggests the formation of a suitable SEI film.[29] Subsequently, the cell is cooled down to 50 °C and cycled with a relevant response in terms stability and efficiency. Indeed, Fig. 7a shows the two plateaus with average voltage of 3.5 and 4 V previously described, and associated to the redox couples $Fe^{2+}/Fe^{3+}$ and $Mn^{2+}/Mn^{3+}$, respectively, with a high reversibility both at the initial stage of the test and after about 100 charge/discharge cycles. It is worth mentioning that this excellent behavior is remarkable, in particular taking into account the test conditions (50 °C in a polymeric matrix) and the operating voltage of the cell (extending over 4.4 V). Furthermore, Fig. 7a reveals a more significant slope of the cell with respect to the cycling test performed using the liquid electrolyte at C/5 within the same voltage and reported in Figure S6 in Supporting Information. This difference may be reasonably ascribed to kinetically limited lithium ions diffusion, both into the polymer matrix and at the electrode/electrolyte interphase.[83] Fig. 7b evidences for the polymer battery an initial capacity of about 110 mAh g$^{-1}$, that is a higher value compared to that of the cell cycling in the liquid electrolyte with the same C-rate and voltage limits (compare Fig. 7b and Fig. S6b in Supporting Information), due to the higher temperature which typically boosts the reaction kinetics in the olivine cathodes.[22,23] The polymer cell reveals a coulombic efficiency of about 90% during the first cycle, which increases to 97 % after the second cycle, and approaches 100% at the final stages of the test (Fig. 7b), with a capacity retention as high as 95 % in 100 cycles. This relevant stability is ascribed to an optimal electrode/electrolyte interphase which slightly modifies during cycles, as demonstrated by the EIS measurement before (Fig. 7c) and after (Fig. 7d) 107 cycles, and by the related NLLS analysis reported in Table 3. The Nyquist plots can be represented by the equivalent circuit $R_e(R_iQ_i)(R_wQ_w)Q_g$, where $R_e$ is the high-frequency intercept accounting for the polymer electrolyte resistance and possible contribution of heterogeneity, $(R_iQ_i)$ represents the middle-high frequency semicircle due to the SEI films and residual charge transfer, $(R_wQ_w)$ is a low-frequency element associated with the Warburg-type Li$^+$ ion diffusion across the polymer, and $Q_g$ is the constant phase element of the cell geometrical capacity.[74,75] The data of Table 3 reveal an overall resistance (R = $R_e$ + $R_i$) with a value increasing from about 330 to about 570, that is, a relatively limited change suitable for allowing an efficient and stable operation of the lithium-polymer battery.[84]

**Figure 7**

**Conclusions**

This paper reported the synthesis and the full characterization of a cathode material exploiting the olivine structure, and benefitting from the simultaneous presence of iron and manganese into a LiFe$_{0.6}$Mn$_{0.4}$PO$_4$ stoichiometry to achieve higher energy compared to the conventional LiFePO$_4$ due to the higher working voltage (see comparison in Figure S7 in Supporting Information).The data indicated the achievement of a material with a well-defined crystalline structure without any impurity due to a process originally optimized by TGA/XRD investigation during the course of the various thermal steps. A suitable micrometric morphology, the expected stoichiometry, and the advantageous presence of 5 wt.% of carbon allowed a reversible electrochemical process of the Fe$^{+2}$/Fe$^{+3}$ and Mn$^{+2}$/Mn$^{+3}$ redox couples in lithium cell at 3.5 and 4 V vs Li$^+$/Li, respectively, with a modest polarization and a relatively low impedance. Furthermore, tests in lithium battery exploiting the conventional electrolyte indicated a maximum capacity of about 130 mAh g$^{-1}$ and an acceptable rate capability from C/10 to 2C rate. Considering the achieved capacity value and the average working voltage of the cells, we can estimate for the LFMP material a theoretical energy density of about 490 Wh kg$^{-1}$. Therefore, taking into account inactive elements such as battery case, current collectors and electronic components, the LFMP can achieve a practical energy approaching 170 Wh kg$^{-1}$ in lithium-ion cell, and even a higher value in cells using the lithium metal anode. Therefore, a step forward for achieving a safe version of the lithium metal battery has been herein proposed by exploiting the polymer configuration, using an electrolyte based on solid PEGDME and operating at 50 °C. The cell reveled an exceptional stability with a capacity of about 110 mAh g$^{-1}$ retained for about 95 % over more than 100 charge/discharge cycles. The results of this work suggest the possible achievement of safe, low cost and efficient batteries including a new generation of stable olivine cathodes, solid polymer electrolyte and the high-energy lithium metal.

**Experimental Section**

*Synthesis of LiFe$_{0.6}$Mn$_{0.4}$PO$_4$*

The olivine cathode (3 g) was prepared by a sol-gel synthesis pathway reported elsewhere for different cathodes.[66,76,85] 1.9808 g of lithium dihydrogen phosphate (LiH$_2$PO$_4$ 99 %, Sigma-Aldrich) was dissolved in 15 mL of water (solution A), whereas transition metal acetates (1.9891 g of iron(II) acetate, Fe(CO$_2$CH$_3$)$_2$ ≥ 99.99 %, Sigma-Aldrich; 1.8685 g manganese(II) acetate tetrahydrate, Mn(CO$_2$CH$_3$)$_2$·4H$_2$O ≥ 99 %, Sigma-Aldrich) and 1.0985 g of citric acid (C$_6$H$_8$O$_7$ ≥ 99.5%, Sigma-Aldrich) were dissolved in 30 mL of water (solution B). Solution A was added dropwise to solution B to obtain an aqueous solution containing Li$^+$, transition metal ions (Fe$^{+2}$ and Mn$^{+2}$), and citric acid in a molar ratio of 1:1:0.3, respectively (solution C), in which the stoichiometry of the reagents was set to achieve a LiFe$_{0.6}$Mn$_{0.4}$PO$_4$ material. Solution C was heated at about 60 °C for ca. 8 hours in a silicon oil (Sigma-Aldrich) bath under stirring through a hot plate until the formation of a dark gel. The gel was initially dried for 12 h at 120 °C under a dry air flow, and afterwards heated under an argon flow at 350 °C with a rate of 2 °C min$^{-1}$, held at 350 °C for 3 h, and naturally cooled to room temperature. The precursor was ground in an agate mortar, heated under an argon flow at 600 °C with a rate of 2 °C min$^{-1}$, held at 600 °C for 10 h, heated at 700 °C with a rate of 5 °C min$^{-1}$, held at 700 °C for 3 h, and naturally cooled to the room temperature. The achieved material was ground in an agate mortar and the sample was identified by the acronym LFMP_700 (i.e., the intermediate olivine). The above sample was subsequently heated under an argon flow at 850 °C with a rate of 5 °C min$^{-1}$, held at 850 °C for 1 h, naturally cooled to the room temperature, ground in an agate and indicated by the acronym LFMP_850 (i.e., the final olivine sample). All the thermal steps of the synthesis procedure were performed in a tubular furnace (GHA 12/3000, Carbolite).

*Composite polymer electrolyte*

The composite polymer electrolyte (CPE) was achieved by using solid polyethylene glycol dimethyl ether (PEGDME2000, $CH_3O(C_2H_4O)_nCH_3$, average MW of 2000 g mol$^{-1}$, Sigma-Aldrich), lithium bis(trifluoromethanesulfonyl)imide (LiTFSI, 99.95% trace metals basis, Sigma-Aldrich), lithium nitrate ($LiNO_3$, 99.99% trace metals basis, Sigma-Aldrich), and fumed silica ($SiO_2$, average particle size 0.007 µm, Sigma-Aldrich) as described in a previous paper.[29] LiTFSI and $LiNO_3$ were mixed with PEGDME2000 in the ratio of 1 mol of each salt to 1 kg of polymer, whilst of $SiO_2$ was included by a weight ratio of 10 % with respect to the mass of PEGDME2000-lithium salts mixture. The electrolyte film was formed through a semi-liquid slurry of the above described powder mixed with acetonitrile (ACN, Sigma-Aldrich), which was subsequently removed upon various drying step to obtain a solid membrane indicated using the acronym PEGDME_CPE.

*Thermogravimetric analysis and SEM/EDS of the olivine cathode*

Thermogravimetric measurement was performed to investigate the synthesis process under a $N_2$ atmosphere employing a heating rate of 2 °C min$^{-1}$ and 5 °C min$^{-1}$ in the 25 – 600 °C and 600 – 850 °C temperature ranges, respectively; while thermogravimetric analyses to check the carbon content of the LFMP_700 and LFMP_850 samples were carried out under air flow employing a heating rate of 5 °C min$^{-1}$ in the 25 – 800 °C temperature range. All measurements were carried out through a Mettler-Toledo TGA 2 instrument. A reaction mechanism (1) involving oxidation of C and Fe in the mixed olivine to form $CO_2$, $Fe_2O_3$, $FePO_4$, $Li_3PO_4$, and $LiMnPO_4$, was assumed for the determination of the carbon content as reported below:

$$xC + yLiFe_{0.6}Mn_{0.4}PO_4 + (0.15y + x)O_2$$
$$\rightarrow 0.1yFe_2O_3 + 0.4yLiMnPO_4 + 0.2yLi_3PO_4 + 0.4yFePO_4 + xCO_2 \quad (1)$$

where *x* and *y* (mols) were determined through the following equations (2):

$$\begin{cases} xM_C + yM_{LiFe_{0.6}Mn_{0.4}PO_4} = A \times m \\ 0.1yM_{Fe_2O_3} + 0.4y + 0.2yM_{Li_3PO_4} + 0.4yM_{FePO_4} = B \times m \end{cases} \quad (2)$$

where A and B are the % in weight achieved from Fig. 2b and Fig. 2d, the values indicated by $M_X$ (g mol$^{-1}$) represent the molecular weight of the various solids involved in reaction (1), and $m$ (g) is the initial weight of sample used for the TGA. It is worth noticing that weight losses before 300 °C were attributed to free or absorbed water removal, those after 300 °C due to $CO_2$ evolution, while weight increases after 200 °C to metal oxide formation.

The carbon amount (wt. %) in the samples was achieved by equation (3):

$$wt.\%C = \frac{xM_C}{xM_C + yM_{LiFe_{0.6}Mn_{0.4}PO_4}} \cdot 100 \quad (3)$$

The scanning electron microscopy (SEM) images of the sample were collected using a Zeiss EVO 40 microscope with a LaB$_6$ thermionic source. Sample stoichiometry was confirmed by energy dispersive X-ray spectroscopy (EDS), collected by the analyzer of the former microscope (X-ACT Cambridge Instrument).

Transmission electron microscopy (TEM) analyses were carried out by a Zeiss EM 910 microscope quipped with a tungsten thermoionic electron gun operating at 100 kV.

*X-ray diffraction (XRD) and Rietveld analysis*

XRD patterns of the sample powder spread on a glass sample holder were collected through a Bruker D8 Advance instrument using a Cu Kα source and a graphite monochromator of the diffracted beam. Scans were performed in the 2θ range from 10° to 90° at a rate of 10 s step$^{-1}$ and a step size of 0.02°. Rietveld refinement of the pattern was carried out through the MAUD software[86] by using the reference parameters of the LiFe$_{0.78}$Mn$_{0.22}$PO$_4$ olivine (*Pnma* space group, N. 62, ICSD 193641).[54] The atomic displacement parameters have been forced to have the same value for Fe, Mn and Li, and for the $PO_3^-$ atoms. The weighted-profile ($R_{wp}$%) and goodness-of-fit ($\sigma$) values were ≤ 20 and ≤ 2.0, respectively (see Table 1).

*Electrodes preparation and cells assembly*

The electrodes were prepared by doctor blade casting of the LFMP_850 (final olivine material) on aluminum carbon coated foils (thickness of 20 µm, MTI Corporation). The slurries were achieved by dispersing in a beaker the active material, poly(vinylidene fluoride) (Solef® 6020 PVDF), and Super P carbon (Timcal) with a 8:1:1 weight ratio, respectively, in *N*-methyl-2-pyrrolidone (Sigma-Aldrich), and stirring at room temperature until homogenization (about 1 hour). The cast electrode foils were dried for 3 h on a hot-plate at 70 °C, cut into the form of disks with a diameter of 10 mm and 14 mm, pressed with a hydraulic press (Silfradent) for 30 s at about 5 tons, and dried for 3 h at 110 °C under vacuum to remove residual traces of water or solvent. The active material loading ranged from 4 to 4.5 mg cm$^{-2}$ as normalized to the electrode geometric area (0.785 and 1.54 cm$^2$ for electrodes with a diameter of 10 mm and 14 mm, respectively). Two-electrode and three-electrode configuration of Swagelok-type cells[87] were assembled by using the composite cathode as the working, lithium-metal disks with diameters of 10 mm as the counter electrode and two 10 mm glass fiber Whatman® GF/A foils as the separator soaked with the electrolyte solution of $LiPF_6$ (1 M) in EC:DMC 1:1 V/V (Sigma-Aldrich battery grade). An additional 10-mm lithium disk was used as and the reference electrode in the 3-electrode configuration cell. CR2032 coin-cells (MTI Corporation) were also assembled by using the cathode disk with a diameter of 14 mm as the working electrode, one 16 mm glass fiber Whatman® GF/B foil the separator soaked with the above described liquid electrolyte, and a lithium disk with a diameter of 14 mm as the counter electrode. All cells were assembled and sealed inside an Ar-filled glovebox (MBraun, $O_2$ and $H_2O$ content lower than 1 ppm).

*Electrochemical tests*

Cyclic voltammetry (CV) measurements were performed on Li|EC:DMC 1:1 V/V, 1M LiPF$_6$|LFMP_850 cell (3-electrode configuration Swagelok) at a scan rate of 0.1 mV s$^{-1}$ in the 2.0 – 4.6 V vs Li$^+$/Li potential range. Electrochemical impedance spectra (EIS) were collected at the open circuit voltage (OCV) condition of the cell, as well as after 1, 5, and 10 CV cycles, and were analyzed through the non-linear least squares (NLLS) fitting method using a Boukamp software (only fits with $\chi^2$ values of the order of 10$^{-4}$ or lower were considered).[74,75] EIS was performed by applying an alternate voltage signal with an amplitude of 10 mV in the frequency range from 500 kHz to 100 mHz. The CV and EIS measurements were carried out by using a VersaSTAT MC Princeton Applied Research (PAR, AMETEK) instrument. Galvanostatic cycles were performed on Li|EC:DMC 1:1 V/V, 1M LiPF$_6$|LFMP_850 cells (2-electrode configuration Swagelok) at the various C-rates in the voltage range of 2.0 – 4.6 V, using an additional constant voltage step at 4.6 V until the current reached ¼ of the nominal C-rate (that is, by adopting the CCCV mode). A rate capability test was performed at increasing current rate by running 5 cycles at C/10, C/5, C/3, C/2, 1C, 2C, and finally lowering back the current to C/10 (1C = 170 mA g$^{-1}$). Furthermore, galvanostatic measurements with a current set at C/10 and C/5 were run for 50 cycles to study the electrode stability in lithium cell. To further characterize the cathode material, a galvanostatic test with the same cell configuration was run at C/5 by restricting the voltage window to 2.4 – 4.4 V with the constant voltage step at 4.4 V. Additional galvanostatic tests aimed to study the long-term cycling behavior of the LFMP_850 material were performed in coin-type lithium cells using the current rate of C/5 and 1C, respectively. All cycling tests were performed using a MACCOR series 4000 battery test system.

*Ex-situ measurements*

After the test at C/5 in coin-type lithium cell, the LFMP_850 electrode was recovered, washed with DMC solvent and dried under vacuum for 30 min, Afterwards, XRD scans of a pristine electrode and the cycled one were collected by a Bruker D8 Advance instrument using a Cu Kα source and a graphite monochromator of the diffracted beam in the 10 – 90° range at a rate of 10 s step$^{-1}$ and a step size of 0.02°, while SEM images of the two samples were acquired through a Zeiss EVO 40 microscope with a $LaB_6$ thermionic source.

*The polymer cell*

A CR2032 coin-cell (MTI Corporation) was assembled by using the LFMP_850 cathode disk with a diameter of 10 mm as the working electrode, the PEGDME_CPE as the separator, two O-rings (CS Hyde, 23-5FEP-2-50) with an internal diameter of 10 mm and thickness of 127 μm for holding the polymer electrolyte, and a lithium disk with diameter of 10 mm as the counter electrode. The Li|PEGDME_CPE|LFMP_850 cell was examined at 50 °C at the constant current rate of C/5 in the voltage range of 2.4 – 4.4 V. EIS measurements were performed on the above cell using an alternate voltage signal with an amplitude of 30 mV in the 500 kHz to 100 mHz frequency range, at the OCV after cell conditioning overnight at 70 °C, at 50°C before cell cycling, and after 107 cycles of the above galvanostatic charge/discharge test. The polymer cell was assembled in Ar-filled glovebox (MBraun, $O_2$ and $H_2O$ content below 1 ppm), cycled using a MACCOR series 4000 battery test system, and subjected to EIS using a VersaSTAT MC Princeton Applied Research (PAR, AMETEK) instrument.

**Acknowledgements**

This work has received funding from the European Union's Horizon 2020 research and innovation programme Graphene Flagship under grant agreement No 881603, and grant "Fondo per l'Incentivazione alla Ricerca (FIR) 2020", University of Ferrara. The authors acknowledge the project "Accordo di Collaborazione Quadro 2015" between University of Ferrara (Department of Chemical and Pharmaceutical Sciences) and Sapienza University of Rome (Department of Chemistry).

**List of Tables**

**Table 1.** Results of Rietveld refinement in terms of lattice parameters, unit cell volume, good-of-fit parameter (GOF), and weighted-profile R factor ($R_{wp}$%) of the intermediate LiFe$_{0.6}$Mn$_{0.4}$PO$_4$ olivine (LFMP_700) and of the final material (LFMP_850). Reference ICSD #193641.

**Table 2.** Equivalent circuits, interphase resistances ($R_1$, $R_2$), and chi-square values accounting for the accuracy ($\chi^2$) of the non-linear least squares (NLLS) analysis[74,75] related to the EIS data of the Li|EC:DMC 1:1 V/V, 1M LiPF$_6$|LFMP_850 three-electrode cell using a Li reference electrode collected at the OCV, and after 1, 5 and 10 CV cycles in the potential range 2.0 – 4.6 V vs. Li$^+$/Li, with a scan rate of 0.1 mV s$^{-1}$. EIS frequency range 500 kHz – 100 mHz; alternate voltage signal amplitude 10 mV. Room temperature (25 °C). See Experimental section for acronyms ad Fig. 4 for relevant voltammetry profiles and Nyquist plots.

**Table 3.** Electrolyte resistance ($R_e$), interphase resistance ($R_i$), and chi-square value indicating the accuracy ($\chi^2$) of the non-linear least squares (NLLS) analysis[74,75] using the equivalent circuit $R_e(R_iQ_i)(R_wQ_w)Q_g$ with the EIS data of the Li|PEGDME_CPE|LFMP_850 cell collected at 50 °C before and after 107 galvanostatic cycles. See the Experimental section for samples' acronyms and Fig. 6(c, d) for relevant Nyquist plots.

**List of Figures**

**Figure 1. (a)** Synthesis process of the LiFe$_{0.6}$Mn$_{0.4}$PO$_4$ olivine cathode evaluated by thermogravimetric analysis (TGA) under nitrogen, with insets depicting the reactions of the citric acid occurring before (left-hand side) and after (right-hand side) the step at 350 °C, and numbers indicating the isotherm steps adopted at various temperatures. **(b-e)** TGA of the isotherm steps taking place at **(b)** 350 °C, **(c)** 600 °C, **(d)** 700 °C, and **(e)** 850 °C. **(f-i)** XRD patterns of **(f)** dry gel, **(g)** precursor at 350°C, **(h)** intermediate sample upon 700 °C (LFMP_700), and **(i)** final electrode material upon 850 °C (LFMP_850). TGA carried out employing heating rates of 2 °C min$^{-1}$ and 5 °C min$^{-1}$ in the 25 – 600 °C and 600 – 850 °C temperature ranges, respectively. XRD patterns collected

performing scans in the 2θ range from 10° to 90° at a rate of 10 s step$^{-1}$ with a step size of 0.02°. See the Experimental section for samples' acronyms.

**Figure 2. (a, c)** Rietveld refinement of the XRD patterns (*Pnma* space group, N. 62) of **(a)** LFMP_700 and **(c)** LFMP_850. In detail: experimental (black dots) and calculated (red line) patterns, difference profile (blue line). **(b, d)** TGA curves of **(b)** LFMP_700 and **(d)** LFMP_850 recorded under air flow in the 25 – 800 °C temperature range at a heating rate of 5 °C min$^{-1}$, and reporting the calculated carbon weight %. See the Experimental section for samples' acronyms and details on carbon % calculation.

**Figure 3.** Scanning electron microscopy (SEM), energy dispersive X-ray spectroscopy (EDS) and transmission electron microscopy (TEM) analyses of the LFMP_850. In detail: **(a, b, c)** SEM images at different magnifications; **(d, e, f, g, h)** SEM-EDS elemental maps of C, Fe, Mn, P, and O, respectively; **(i, j, k)** TEM images at various magnifications. See the Experimental section for the sample's acronym.

**Figure 4. (a)** Cyclic voltammetry (CV) and **(b)** electrochemical impedance spectroscopy (EIS) measurements of the Li|EC:DMC 1:1 V/V, 1M LiPF$_6$|LFMP_850 three-electrode cell with a Li reference electrode. CV potential range 2.0 – 4.6 V vs. Li$^+$/Li; scan rate 0.1 mV s$^{-1}$. EIS carried out at the OCV of the cell and after 1, 5 and 10 voltammetry cycles (insets report magnifications in the middle-high frequency region); frequency range 500 kHz – 100 mHz; alternate voltage signal amplitude 10 mV. See the Experimental section for the sample's acronym.

**Figure 5.** Galvanostatic cycling performances of the the Li|EC:DMC 1:1 V/V, 1M LiPF$_6$|LFMP_850 cell. **(a, b)** Rate capability of in terms of **(a)** voltage profile and **(b)** corresponding cycling trend performed at C/10, C/5, C/3, C/2, 1C and 2C currents (1C = 170 mA g$^{-1}$). **(c-f)** Electrochemical performances of the cell at the constant current rate of **(c, d)** C/10 and **(e, f)** C/5 in terms of **(c, e)** voltage profiles and **(d, f)** cycling trend with discharge capacity in left-hand side *y*-axis and coulombic

efficiency in right-hand side *y*-axis. Voltage range 2.0 – 4.6 V with an additional constant voltage step at 4.6 V (CCCV mode) until a final current of ¼ referred to the nominal C-rate. Room temperature (25 °C). See the Experimental section for sample's acronym.

**Figure 6. (a)** Cycling trend related to a galvanostatic cycling tests performed at C/5 current (1C = 170 mA g$^{-1}$) on a Li|EC:DMC 1:1 V/V, 1M LiPF$_6$|LFMP_850 coin-cell between 2.0 and 4.6 V with an additional constant voltage step at 4.6 V (CCCV mode) until a final current of ¼ referred to the nominal C-rate (inset displays a voltage profile collected at the steady state); **(b)** Comparison between XRD patterns of a pristine LFMP_850 electrode (red) and the LFMP_850 electrode recovered upon the galvanostatic cycling measurement in lithium cell reported in panel a; **(c-f)** SEM images of **(c, d)** the pristine LFMP_850 electrode and **(e, f)** the LFMP_850 electrode recovered upon the cycling test in lithium cell displayed in panel a. Room temperature (25 °C). See the Experimental section for sample's acronym.

**Figure 7.** Electrochemical performances of the Li|PEGDME_CPE|LFMP_850 polymer cell at 50 °C. **(a, b)** Galvanostatic cycling in terms of **(a)** voltage profiles and **(b)** cycling trend with discharge capacity in left-hand side *y*-axis and coulombic efficiency in right-hand side *y*-axis performed at C/5 current rate (1C = 170 mA g$^{-1}$). **(c, d)** Electrochemical impedance spectroscopy (EIS) at **(c)** the OCV condition and **(d)** after 107 galvanostatic cycles at C/5 current rate. Voltage range: 2.4 – 4.4 V. EIS frequency range 500 kHz – 100 mHz, alternate voltage signal amplitude 30 mV. See the Experimental section for the samples' acronyms.

| Sample | a (Å) | b (Å) | c (Å) | V (Å$^3$) | Crystallite size (nm) | GOF ($\sigma$) | Rwp% |
|---|---|---|---|---|---|---|---|
| ICSD #193641 | 10.343 | 6.022 | 4.704 | 293.03 | 100 | / | / |
| LFMP_700 | 10.369 ± 4×10$^{-4}$ | 6.039 ± 2×10$^{-4}$ | 4.709 ± 2×10$^{-4}$ | 294.94 ± 0.03 | 95 ± 2 | 1.33 | 14.54 |
| LFMP_850 | 10.378 ± 3×10$^{-4}$ | 6.044 ± 2×10$^{-4}$ | 4.712 ± 2×10$^{-4}$ | 295.80 ± 0.03 | 147 ± 4 | 1.26 | 13.89 |

**Table 1**

| Cell condition | Equivalent circuit | $R_1$ (Ω) | $R_2$ (Ω) | $R_i = R_1 + R_2$ (Ω) | $\chi^2$ |
|---|---|---|---|---|---|
| OCV | $R_e(R_1Q_1)Q_g$ | 19.5 ± 0.3 | / | 19.5 ± 0.3 | 6 x 10$^{-4}$ |
| After 1 cycle | $R_e(R_1Q_1)(R_2Q_2)Q_g$ | 4.0 ± 0.1 | 11.7 ± 0.2 | 15.7 ± 0.2 | 4 x 10$^{-5}$ |
| After 5 cycles | $R_e(R_1Q_1)(R_2Q_2)Q_g$ | 1.2 ± 0.1 | 7.2 ± 0.1 | 8.4 ± 0.1 | 5 x 10$^{-5}$ |
| After 10 cycles | $R_e(R_1Q_1)(R_2Q_2)Q_g$ | 1.4 ± 0.1 | 6.1 ± 0.1 | 7.5 ± 0.1 | 2 x 10$^{-5}$ |

**Table 2**

| Cell condition | $R_e$ (Ω) | $R_i$ (Ω) | $R = R_e + R_i$ (Ω) | $\chi^2$ |
|---|---|---|---|---|
| OCV | 272 ± 1 | 58 ± 3 | 330 ± 3 | $2 \times 10^{-5}$ |
| After 107 cycles | 194 ± 34 | 273 ± 42 | 571 ± 54 | $2 \times 10^{-4}$ |

**Table 3**

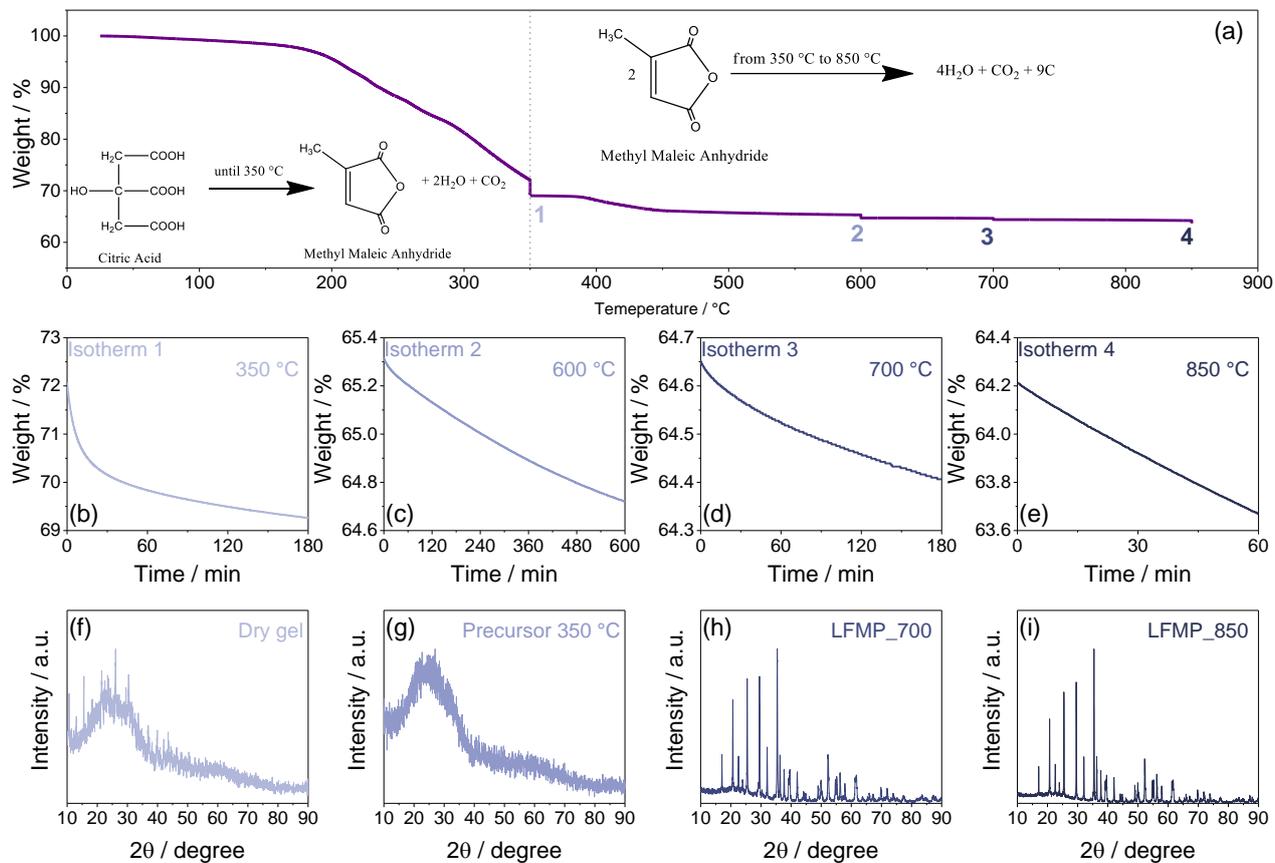

**Figure 1**

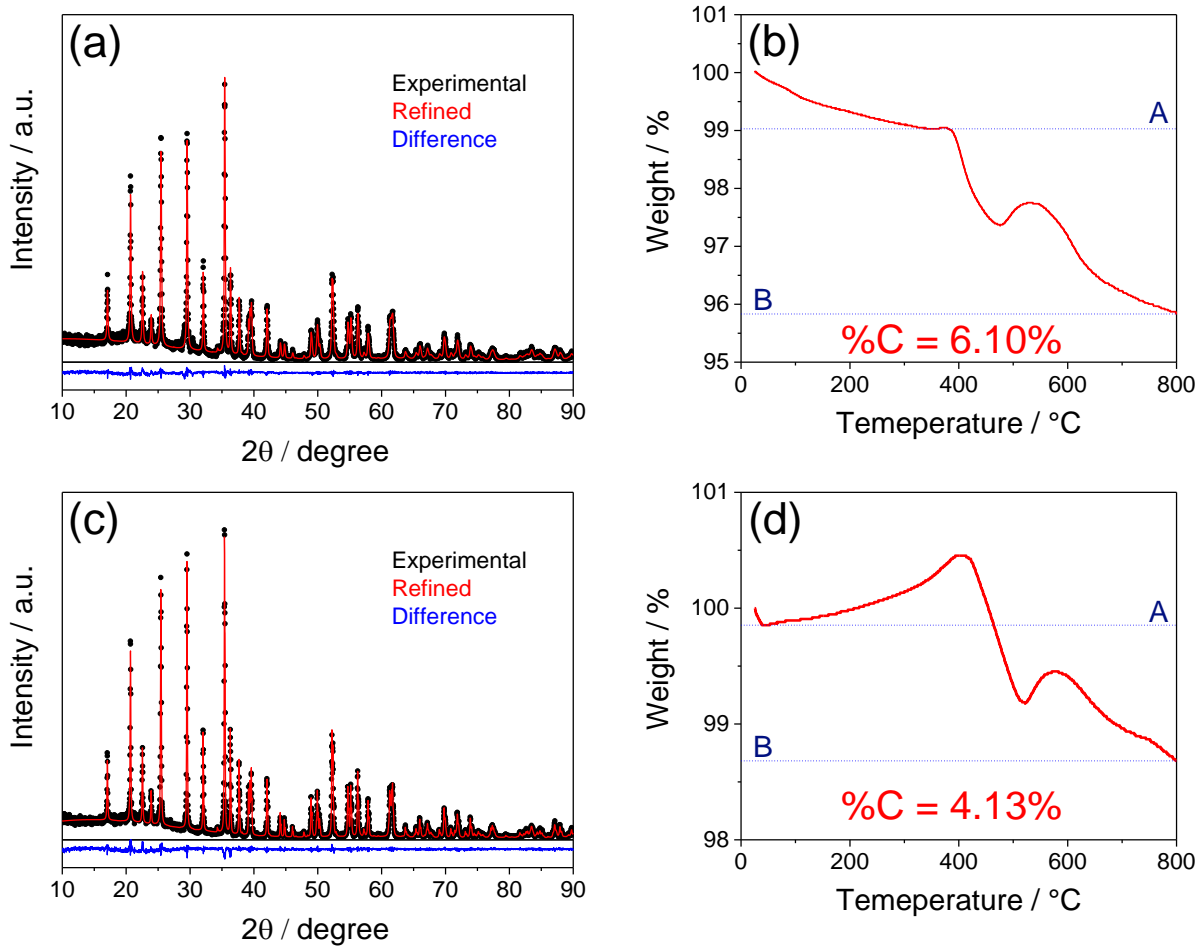

**Figure 2**

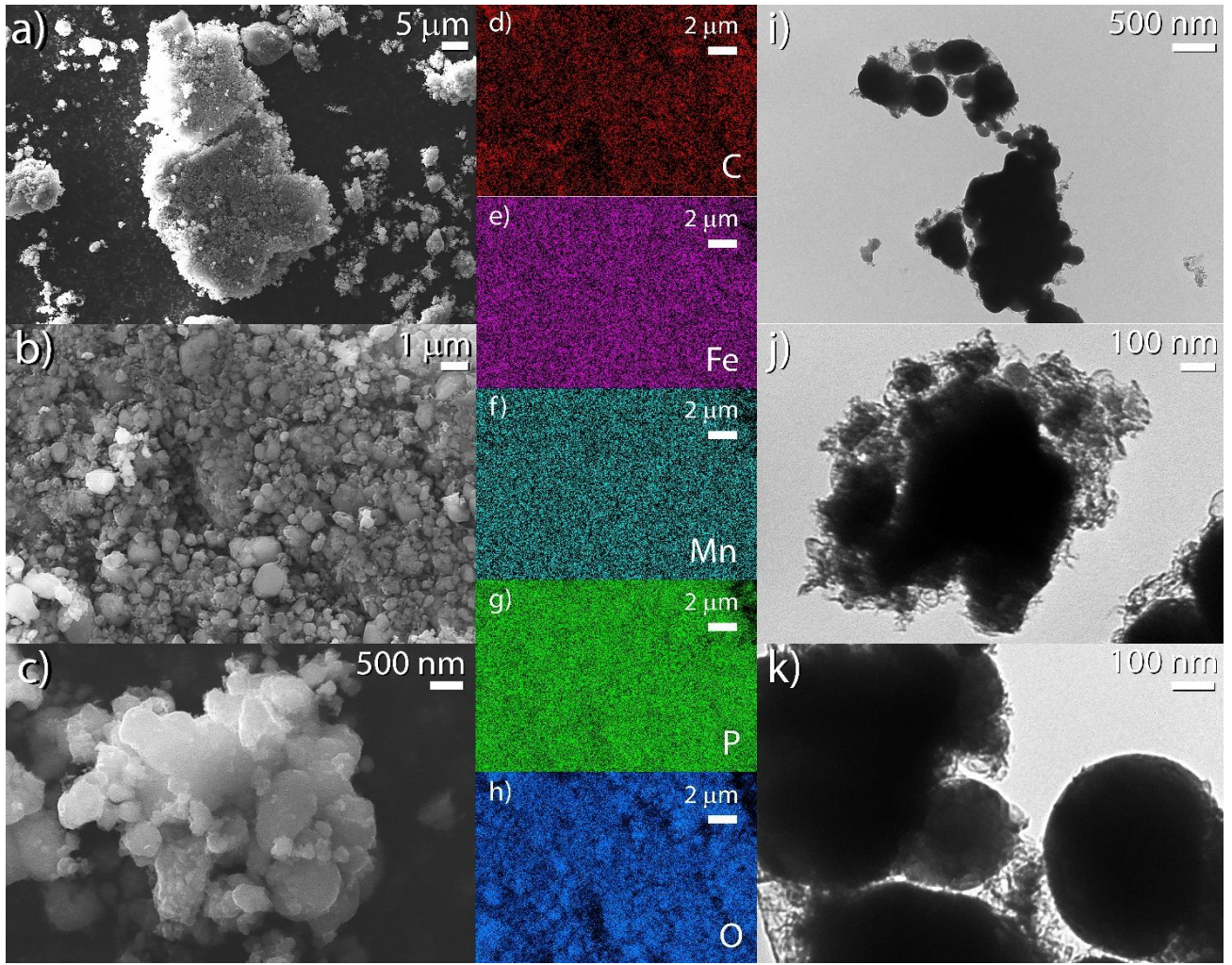

**Figure 3**

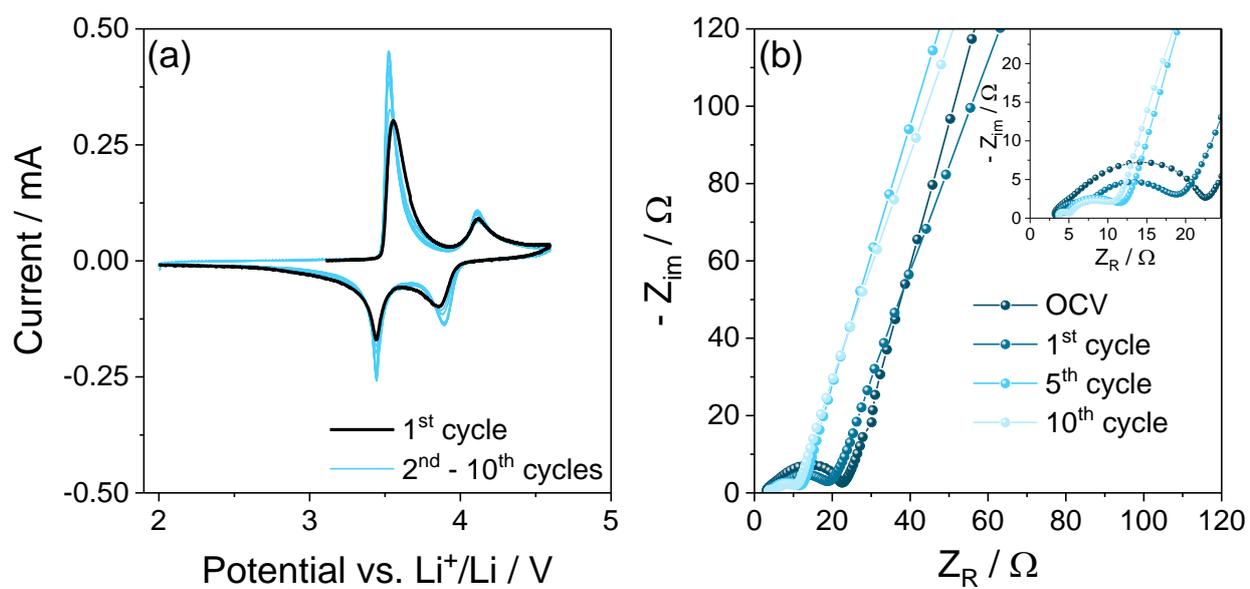

**Figure 4**

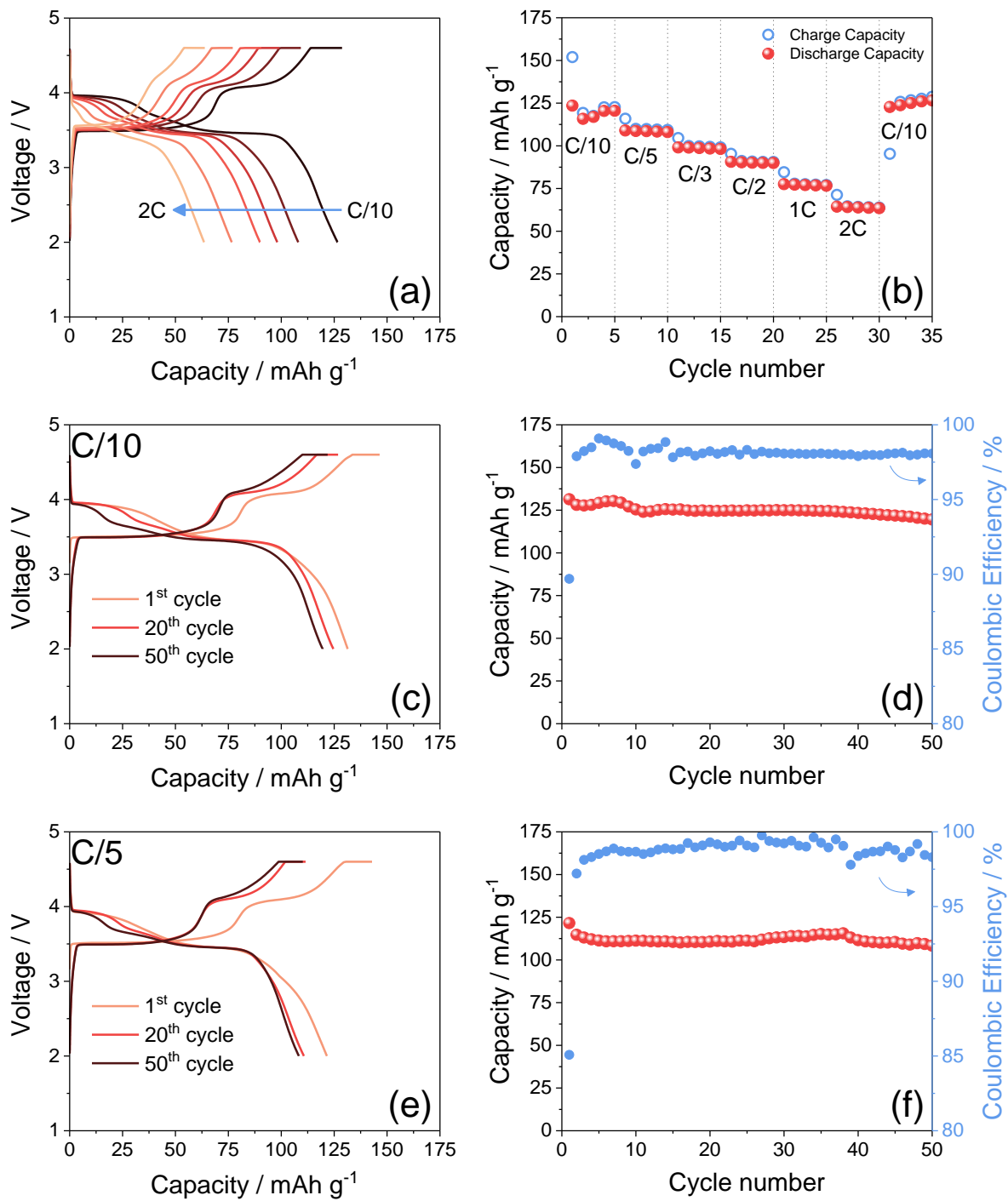

**Figure 5**

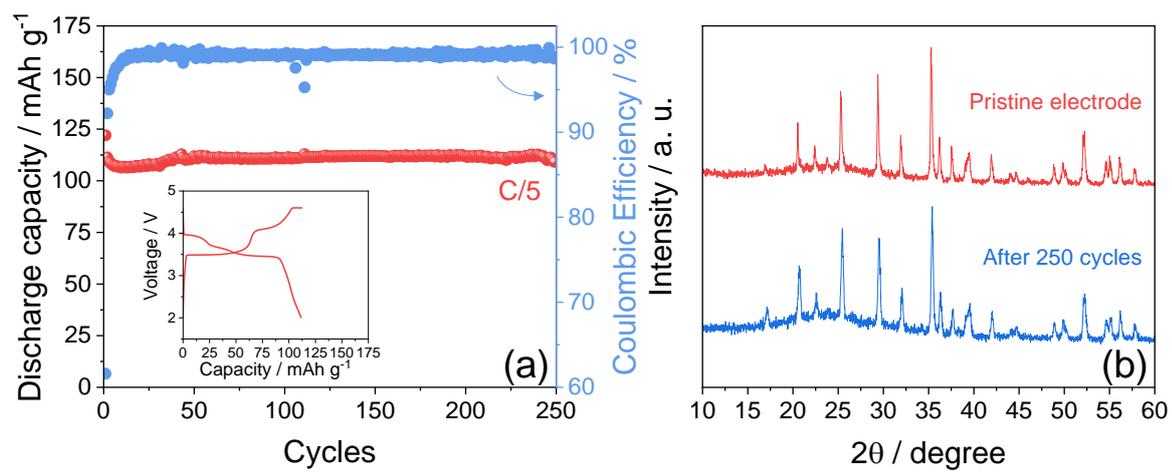
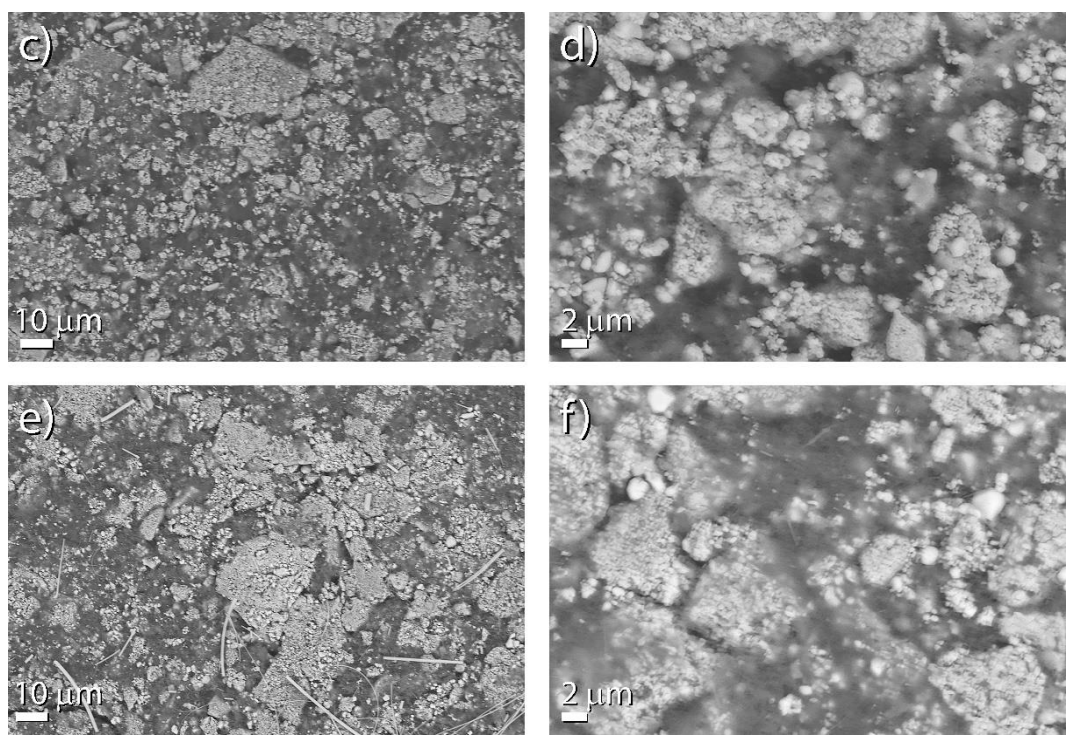

**Figure 6**

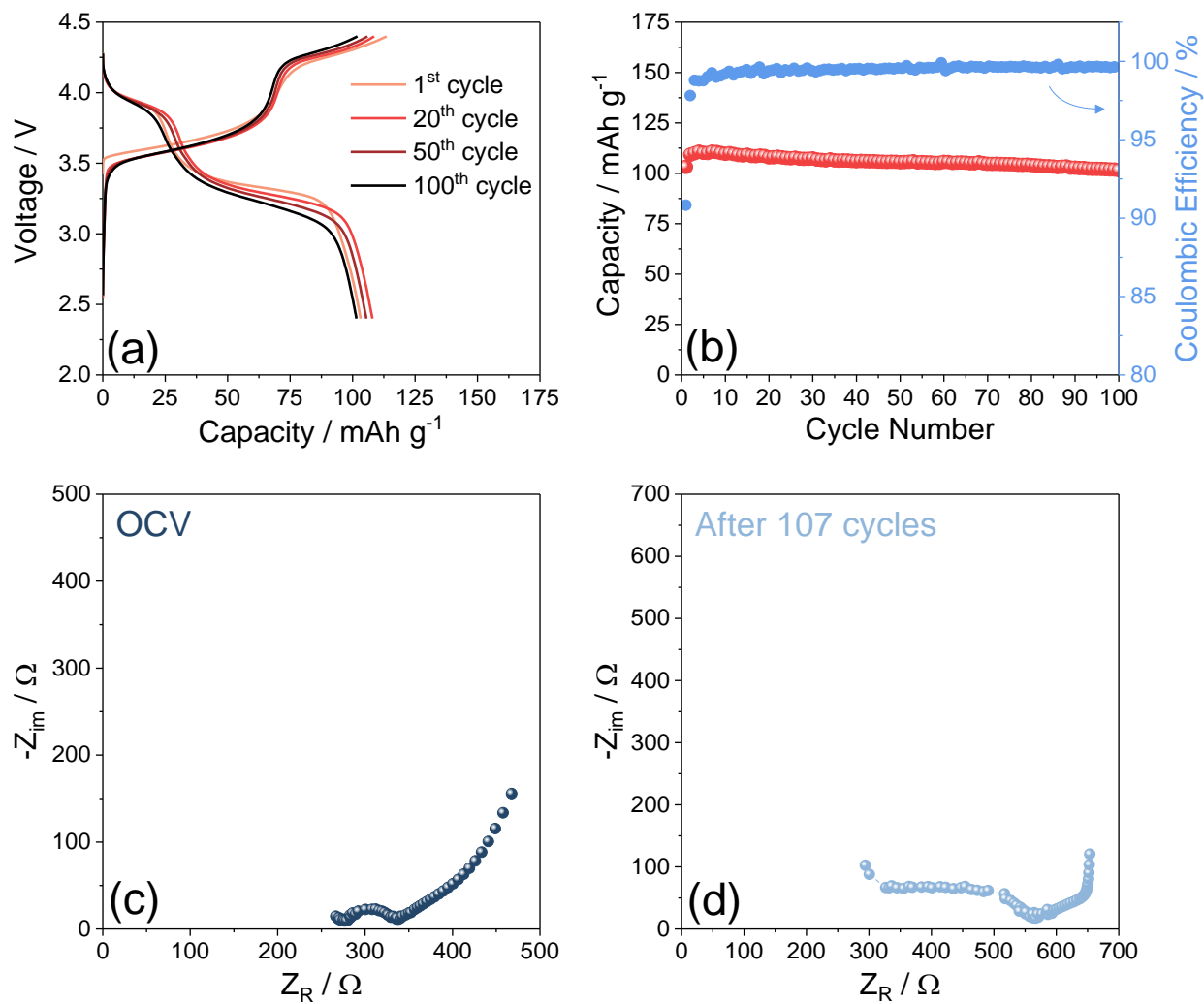

**Figure 7**

**Synthesis and Characterization of a LiFe$_{0.6}$Mn$_{0.4}$PO$_4$ Olivine Cathode for Application in a New Lithium Polymer Battery**

Luca Minnetti[1], Vittorio Marangon[1], and Jusef Hassoun[1,2,3,*]

[1] *Graphene Labs, Istituto Italiano di Tecnologia, via Morego 30, Genova, 16163, Italy*

[2] *University of Ferrara, Department of Chemical, Pharmaceutical and Agricultural Sciences, Via Fossato di Mortara 17, 44121, Ferrara, Italy.*

[3] *National Interuniversity Consortium of Materials Science and Technology (INSTM) University of Ferrara Research Unit, University of Ferrara, Via Fossato di Mortara, 17, 44121, Ferrara, Italy.*

Corresponding Author: jusef.hassoun@iit.it, jusef.hassoun@unife.it

**Keywords**

Olivine cathode; LiFe$_{0.6}$Mn$_{0.4}$PO$_4$; Lithium metal anode; Polymer electrolyte; PEGDME

**Supporting Information**

Figure S1 reports the XRD patterns of LFMP_700, LFMP_850 (see experimental section of the manuscript for acronym identification), the reference diffractogram (ICSD #193641),[1] and shows the peaks indexing (Fig. S1a). In addition, Fig. S1b and Fig. S1c depict the gaussian fitting of the main peak (2θ = 35.4°) of LFMP_700 and LFMP_850, respectively, in order to obtain the full width half maximum (FWHM) and estimate the average size of the crystallite (D, Å) by using the Scherrer equation (1):[2]

$$D = \frac{k\lambda}{\beta \cos\theta} \qquad (1)$$

where λ is the wavelength of the X-ray radiations (1.5418 Å for the Cu Kα radiation used herein), k is a dimensionless shape factor, with a value about 0.9, θ is diffraction angle of the main peak and β is the full width half maximum calculated by equation (2):

$$\beta = \frac{22 * FWHM}{180 * 7} \qquad (2)$$

The olivine structure of the intermediate sample (LFMP_700) and the final one (LFMP_850) is confirmed by the peak indexing (Fig. S1a), and the analysis suggests an increase of D from a value of about 405 Å for LFMP_700 (Fig. S1b) to about 501 Å for LFMP_850 (Fig. S1c), most likely due to partial coalescence of the crystalline domains during the last thermal step. We remark that the achievement of a suitable structure and the absence of excessive deformations suggest the applicability of the electrode in lithium-ion batteries[3] (For further details on the material structure see Fig. 2a, Fig. 2c and Tab. 1 in the manuscript).

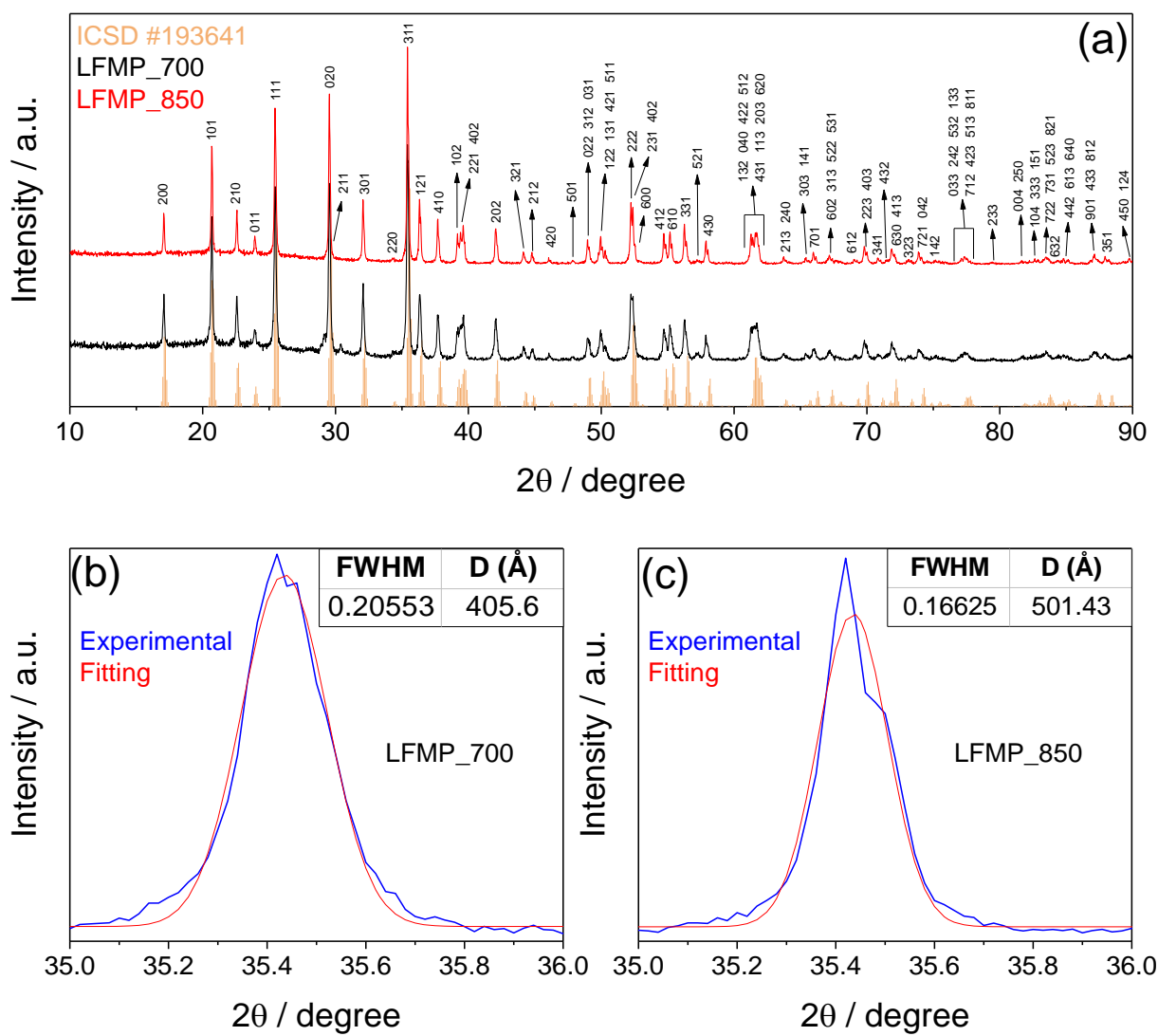

**Figure S1. (a)** XRD patterns of LFMP_700 and LFMP_850 indexed according to the peaks of the reference diffractogram (ICSD #193641).[1] **(b, c)** Main peak at 35.4° of 2θ (blue plot) and fitted one (red plot) for the application of Scherrer equation,[2] including full width half maximum (FWHM) and the average size of the crystallite (D) for **(b)** LFMP_700 and **(c)** LFMP_850. 2θ scan from 10° to 90° at a rate of 10 s step$^{-1}$ with a step size of 0.02°. See Experimental section in the manuscript for the samples' acronyms.

Figure S2 reports the analysis of the SEM image of the LFMP_850 aimed to determine the particle size distribution. The image reveals a maximum in the distribution curve centered at about 500 nm, which is in line with the SEM images in Fig. 3 in the manuscript that indicates the presence of agglomerates formed by primary particles with a submicron diameter ranging from 300 to 900 nm.

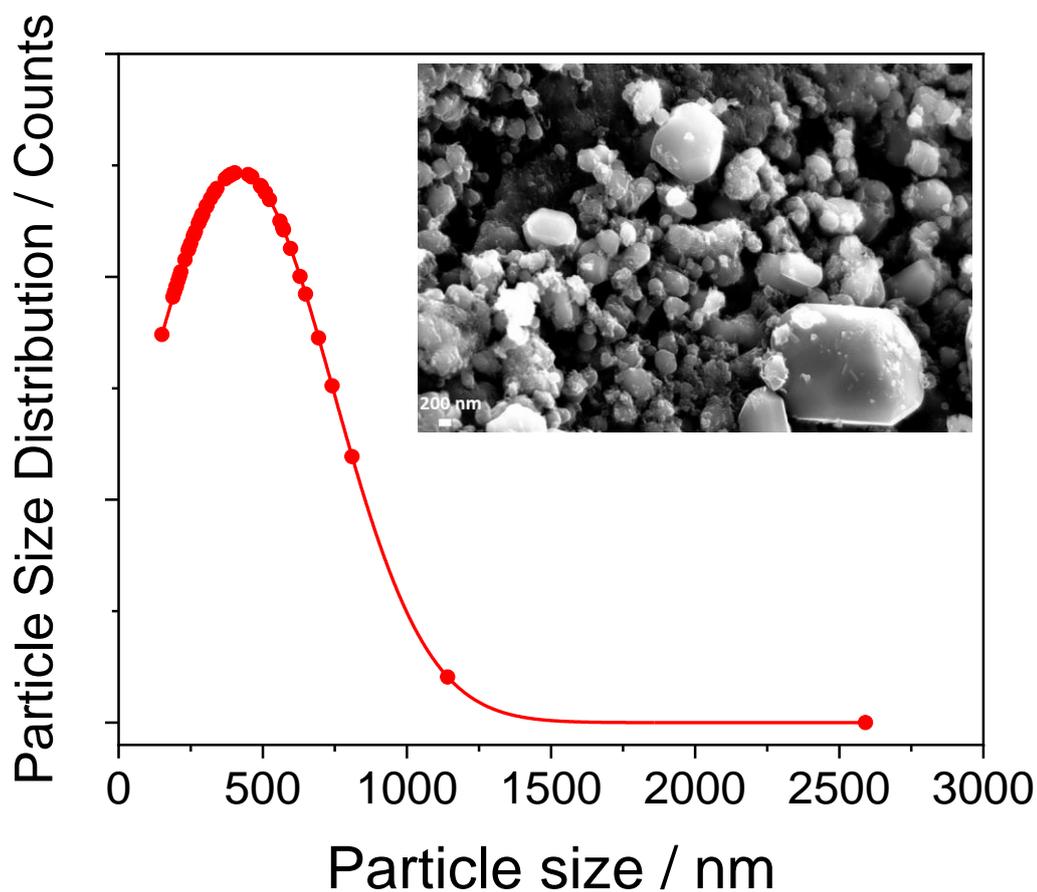

**Figure S2**. Particle size distribution of the LFMP_850 determined by the analysis of the SEM image reported in inset. See Experimental Section in the manuscript for the sample's acronym.

Figure S3 shows the EDS spectrum of LFMP_850 collected to evaluate the amount of Fe, Mn and P in the material. The results confirm in a first approximation the stoichiometry expected by the formula LiFe$_{0.6}$Mn$_{0.4}$PO$_4$, with molar ratios for Fe and Mn of 0.57 and 0.43, respectively. Further morphological investigation including EDS mapping is shown in Fig. 3 of the manuscript.

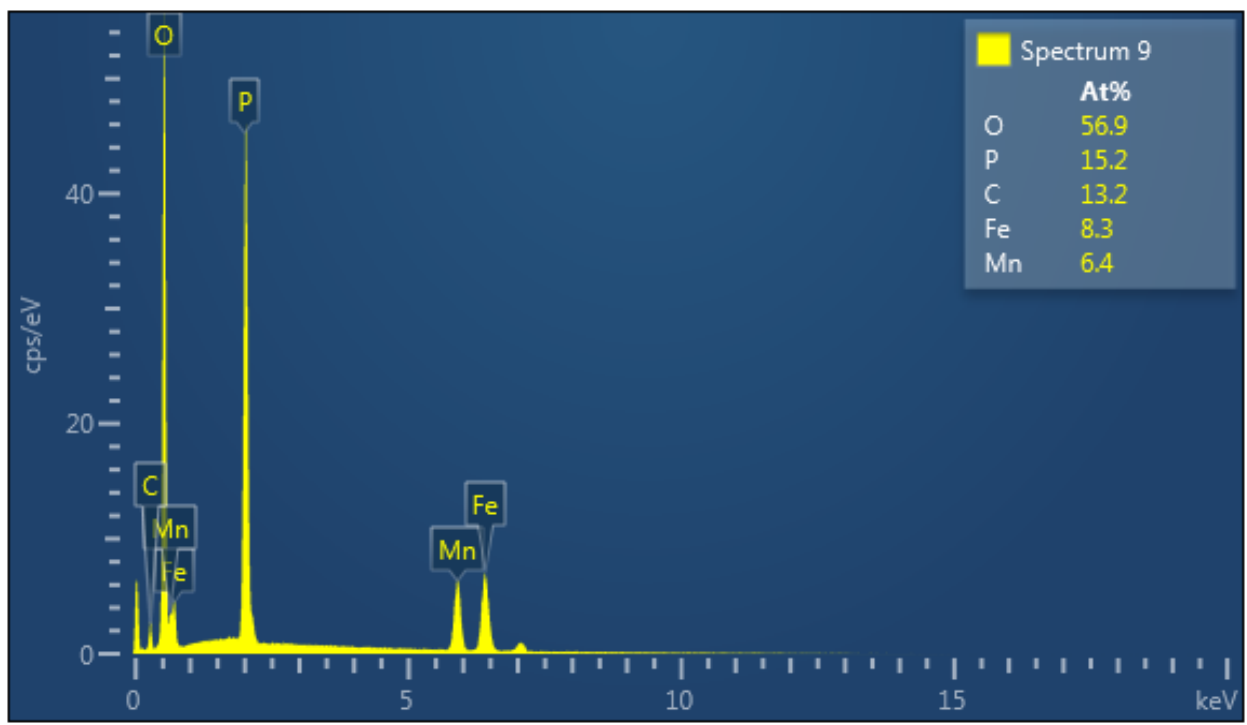

**Figure S3.** EDS spectrum of LFMP_850 olivine cathode. See Experimental Section in the manuscript for the sample's acronym.

Figure S4 shows the cycling tests prolonged over 300 charge/discharge of the LFMP_850 in lithium coin-cells at 1C in terms of the steady state voltage profile (panel a) and capacity with efficiency trend (panel b). The figure reveals a remarkable stability with capacity retention exceeding 85 %, thus accounting for the optimal structure and morphology of the LFMP_850 which is reflected into a well-reversible Li-(de)insertion process in battery operating at the room temperature with the conventional liquid electrolyte (i.e., EC:DMC 1:1 V/V, LiPF$_6$ 1M).

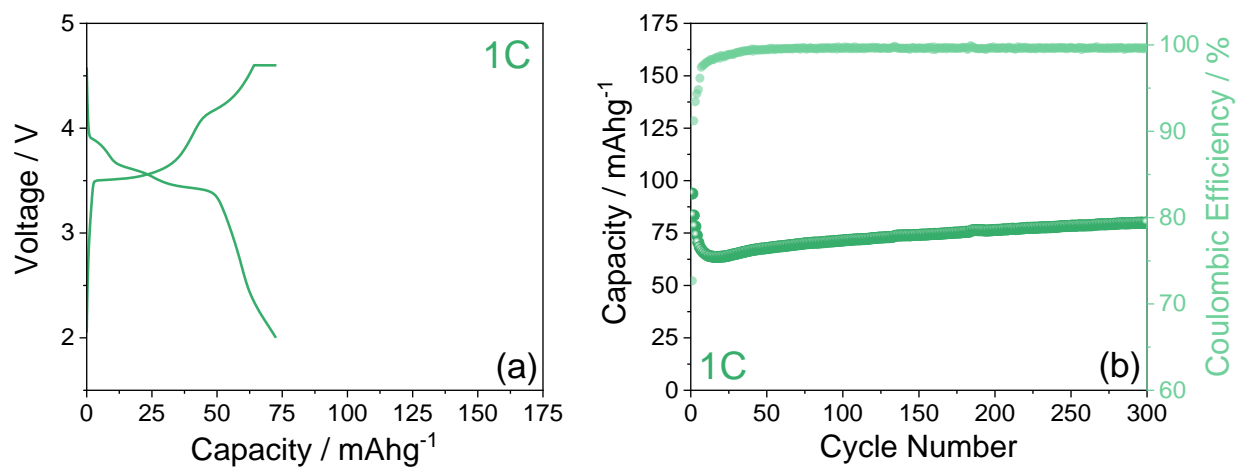

**Figure S4.** Galvanostatic performances of the Li|EC:DMC 1:1 V/V, 1M LiPF$_6$|LFMP_850 cell at the constant current rate of 1C in terms of **(a)** steady state voltage profile and **(b)** prolonged cycling trend with discharge capacity in left-hand side *y*-axis and coulombic efficiency in right-hand side *y*-axis (1C = 170 mA g$^{-1}$). Voltage range 2.0 – 4.6 V with an additional constant voltage step at 4.6 V (CCCV mode) until a final current of ¼ referred to the nominal C-rate. Room temperature (25 °C). See the Experimental section for sample's acronym.

Figure S5 shows the Nyquist plot of the EIS measurement carried out on the Li|PEGDME_CPE|LFMP_850 cell upon aging overnight at 70 °C at the OCV (see experimental section of the manuscript for acronyms). A NLLS analysis of the spectrum is performed using the equivalent circuit R$_e$(R$_i$Q$_i$)(R$_w$Q$_w$)Q$_g$, and the results are reported in Table S1 in terms of electrolyte resistance (R$_e$), interphase resistance (R$_i$), and χ$^2$ parameter. The interphase resistance of the polymer cell at 70 °C, represented by the width of the high-middle frequency semicircle of the Nyquist plot, incorporates different contributes including SEI film, possible charge transfer and heterogeneity.[4] The total resistance of about 275 Ω of the cell at 70 °C, including the electrolyte and the interphase, appears smaller than that of that at the same condition (OCV) cooled down to 50 °C (330 Ω in Tab. 3 of the manuscript), likely due to the a higher ionic conductivity of the PEGDME_CPE solid electrolyte.[5]

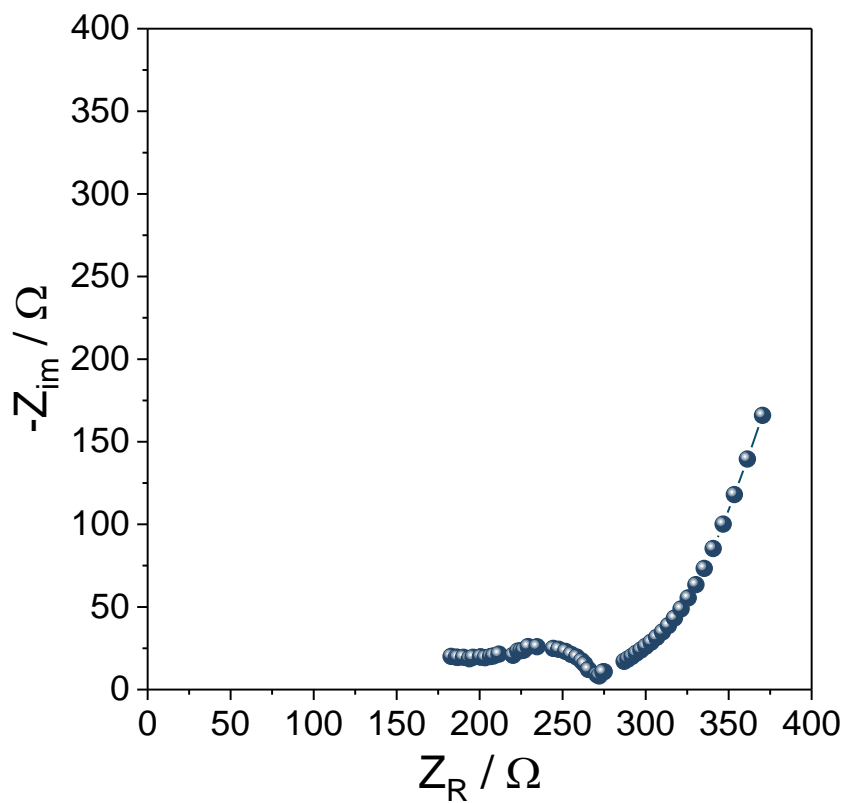

**Figure S5.** Electrochemical impedance spectroscopy (EIS) Nyquist plot of the Li|PEGDME_CPE|LFMP_850 cell collected at 70 °C at the OCV upon aging overnight. Frequency range 500 kHz – 100 mHz; alternate voltage signal amplitude 30 mV. See the Experimental Section in the manuscript for the samples' acronyms.

| Cell condition | $R_e$ (Ω) | $R_i$ (Ω) | R = $R_e$ + $R_i$ (Ω) | $\chi^2$ |
|---|---|---|---|---|
| OCV | 194 ± 3 | 81 ± 5 | 275 ± 6 | 4 x 10$^{-5}$ |

**Table S1.** Interphase resistance ($R_1$), electrolyte resistance ($R_e$), and chi-square value indicating the accuracy ($\chi^2$) of the non-linear least squares (NLLS) analysis[6,7] using the equivalent circuit $R_e(R_iQ_i)(R_wQ_w)Q_g$ and the EIS Nyquist plot in Fig. S4 of the Li|PEGDME_CPE|LFMP_850 cell upon aging overnight at 70 °C in the OCV condition. See Experimental Section in the manuscript for the samples' acronyms.

Figure S6 shows the room temperature electrochemical performances of the Li|EC:DMC 1:1 V/V, 1M LiPF$_6$|LFMP_850 cell at C/5 in terms of voltage profiles (panel a) and cycling behaviour with efficiency (panel b) in the voltage window restricted to 2.4 – 4.4 V. The cell shows a capacity of about 95 mAh g$^{-1}$ with a coulombic efficiency of about 81% during the first cycle, increasing to about 100% upon cycling, and a retention of 95 % over 100 cycles, thus confirming the high stability of the olivine cathode. The test is performed for comparison with the Li|PEGDME_CPE|LFMP_850 cell using the same voltage range reported in Fig. 6 in the manuscript. The data indicate comparable stability and efficiency for the two cells, as well as a higher capacity for the latter compared to the former (i.e., of about 110 mAh g$^{-1}$ and 95 mAh g$^{-1}$, respectively) as ascribed to the higher operating temperature which improves the olivine cathode kinetics.[8]

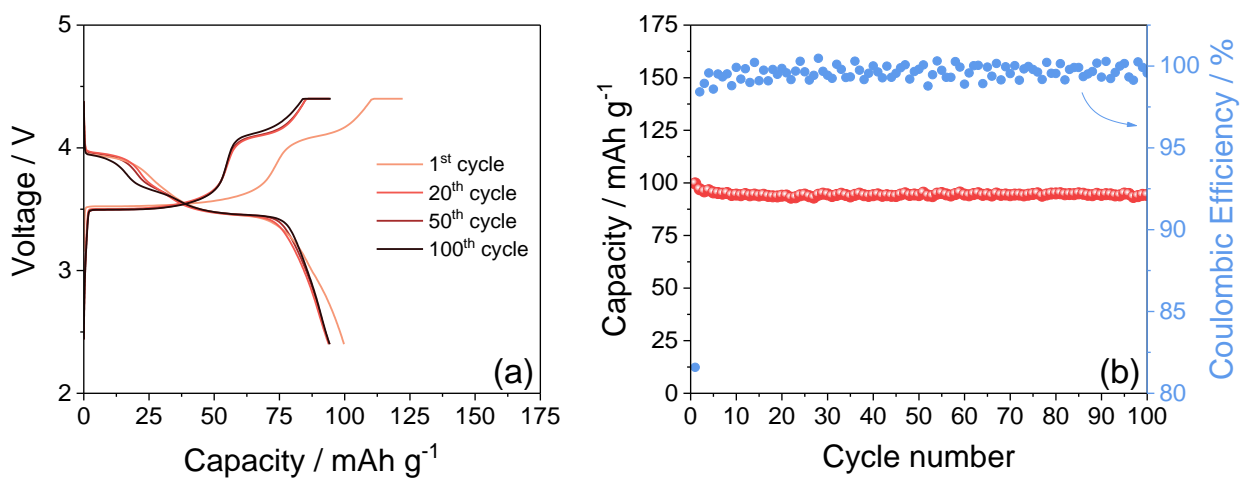

**Figure S6.** Galvanostatic performance of the Li|EC:DMC 1:1 V/V, 1M LiPF$_6$|LFMP_850 cell at C/5 current rate (1C = 170 mA g$^{-1}$) in terms of **(a)** steady state voltage profiles and **(b)** cycling trend with discharge capacity in left-hand side *y*-axis and coulombic efficiency in right-hand side *y*-axis. Voltage range 2.4 – 4.4 V with an additional constant voltage step at 4.4 V (CCCV mode) until a final current of ¼ referred to the nominal C-rate. Room temperature (25 °C). See the Experimental section for sample's acronym.

Figure S7 compares the performance of LMFP_850 and commercial LFP at C/10 in terms of voltage profile. The two materials reveal very similar cycle life and efficiency (data not reported), and Figure S6 evidences that LFP has a higher specific capacity compared to LMFP_850, that is, 150 mAh g$^{-1}$ and 130 mAh g$^{-1}$, respectively. However, the former reveals working voltage of about 3.75 which is higher than that ascribed to LFP (i.e., of about 3.45 V). Hence, we believe that the LMFP_850 sample may achieve a higher energy density compared to LFP upon further optimization of the synthesis conditions.

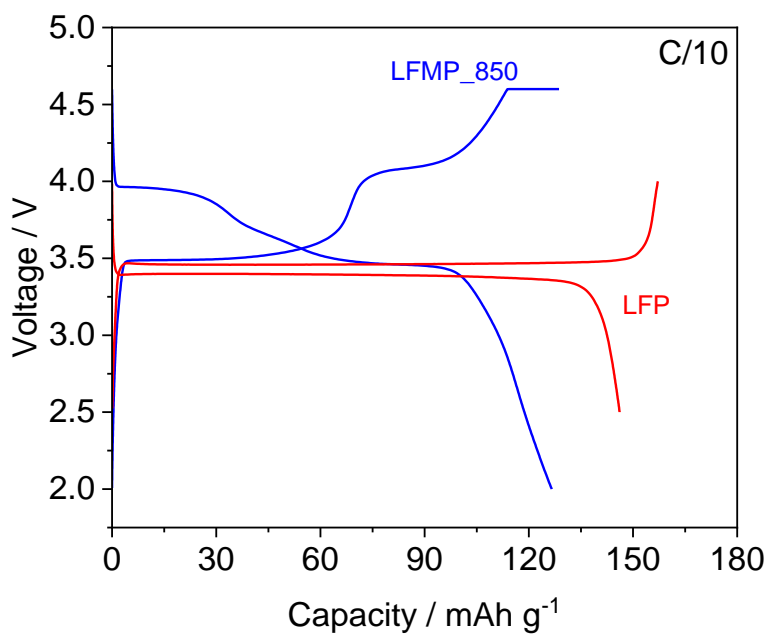

**Figure S7**. Steady state voltage profile of the galvanostatic cycling of the Li|EC:DMC 1:1 V/V, 1M LiPF$_6$|Cathode cell at C/10 current rate (1C = 170 mA g$^{-1}$) where the cathode is either LFP (red) or LFMP_850 (blue). Voltage range 2.5 – 4.0 for LFP and 2.4 – 4.4 V with an additional constant voltage step at 4.4 V (CCCV mode) until a final current of ¼ referred to the nominal C-rate for LFMP_850. Room temperature (25 °C). See the Experimental section for sample's acronym.